\documentclass[prd,twocolumn,showpacs,nofootinbib]{revtex4}
\usepackage{amsmath}
\usepackage{txfonts}

\usepackage{xspace}
\usepackage{multirow}
\usepackage{rotating}
\usepackage{amssymb}
\usepackage{myaasmacros}
\usepackage{comment}
\usepackage{color}

\newcommand{\dd}{\mathrm{d}}

\newcommand{\nHI}{n_{\mathrm{HI}}}

\newcommand{\sigHI}{\sigma_{\mathrm{HI}}}
\newcommand{\sigHIbar}{\bar{\sigma}_{\mathrm{HI}}}
\newcommand{\kappat}{\kappa_{\mathrm{tot}}}
\newcommand{\kappatb}{\kappa_{\mathrm{tot},0}}
\newcommand{\kappaHI}{\kappa_{\mathrm{HI}}}

\newcommand{\betaHI}{\beta_{\mathrm{HI}}}
\newcommand{\betaz}{\beta_z}
\newcommand{\betaV}{\beta_V}
\newcommand{\bclump}{b_{\mathrm{clump}}}
\newcommand{\bHI}{b_{\mathrm{HI}}}
\newcommand{\frec}{f_{\mathrm{rec}}}
\newcommand{\betarec}{\beta_{\mathrm{r}}}
\newcommand{\betar}{\betarec}
\newcommand{\fLL}{f_{\mathrm{LL}}}
\newcommand{\nuLL}{\nu_{\mathrm{LL}}}

\newcommand{\Gammab}{\Gamma_0}
\newcommand{\Mpc}{\mathrm{Mpc}}
\newcommand{\HI}{H{\sc i}\xspace}
\newcommand{\hi}{\HI}
\newcommand{\cm}{\mathrm{cm}}
\newcommand{\bjeff}{b_{j,\mathrm{eff}}}
\newcommand{\qjeff}{q_{j,\mathrm{eff}}}
\newcommand{\qHI}{q_{\mathrm{HI}}}
\newcommand{\km}{\mathrm{km}}
\newcommand{\deltaSN}{\delta_{\mathrm{SN}}}

\newcommand{\kappaclump}{\kappa_{\mathrm{clump}}}
\newcommand{\betaclump}{\beta_{\mathrm{clump}}}
\newcommand{\pclump}{p_{\mathrm{clump}}}
\newcommand{\Msol}{\mathrm{M}_{\odot}}
\renewcommand{\vec}[1]{\mathbf{#1}}
\newcommand{\kappaclumpbar}{\bar{\kappa}_{\mathrm{clump}}}
\newcommand{\deltau}{\updated{\delta_{\nHI,\mathrm{u}}}}
\newcommand{\bu}{\updated{b_{\mathrm{HI,u}}}}
\newcommand{\tdelta}{\hspace{0.5mm}\tilde{\hspace{-0.5mm}\delta}}

\newcommand{\tdeltau}{\updated{\tdelta_{\nHI,\mathrm{u}}}}
\newcommand{\tdeltaHI}{\tdelta_{\nHI}}
\newcommand{\tdeltaSN}{\tdelta_{\mathrm{SN}}}
\newcommand{\updated}[1]{#1}
\begin{document}
 
\title{Scale-dependent bias in the BAO-scale intergalactic neutral hydrogen}

\author{Andrew Pontzen}
\email{a.pontzen@ucl.ac.uk}
\address{Department of Physics and Astronomy, University College
  London, Gower Street, London WC1E 6BT, UK}
\date{3 February 2014}

\begin{abstract}
  I discuss fluctuations in the neutral hydrogen density of the $z
  \approx 2.3$ intergalactic medium and show that their relation to
  cosmic overdensity is strongly scale-dependent. This behaviour arises
  from a linearized version of the well-known ``proximity effect'', in
  which bright sources suppress atomic hydrogen density. Using a
  novel, systematic and detailed linear-theory radiative transfer
  calculation, I demonstrate how \hi density consequently
  anti-correlates with total matter density when averaged on scales
  exceeding the Lyman-limit mean-free-path.

  The radiative transfer thumbprint is highly distinctive and should
  be measurable in the Lyman-$\alpha$ forest. Effects extend to
  sufficiently small scales to generate significant distortion of the
  correlation function shape around the baryon acoustic oscillation
  peak, although the peak location shifts only by $1.2$ percent
  for a mean source bias of $b_j=3$. The distortion changes significantly
  with $b_j$ and other astrophysical parameters; measuring it should
  provide a helpful observational constraint on the nature of ionizing
  photon sources in the near future.
\end{abstract}


\maketitle

\section{Introduction}

The Lyman-$\alpha$ forest \cite{Rauch98ARAA} is the imprint of the
intergalactic medium (IGM) -- specifically, neutral hydrogen -- on the
spectra of distant quasars.  At high redshift the rapidly-changing
forest probes hydrogen reionization
\cite{Becker01Reionization,Fan02Reionization,2006ARA&A..44..415F}; at
lower redshift, the forest has a steadier ionization state and is used
to trace overall matter density fluctuations
\cite{Croft02,McDonald03,BOSS_Slosar11}. Correlating Lyman-$\alpha$
fluctuations over small scales therefore places a strong constraint on
modifications to the standard cold dark matter picture of structure
formation
\cite{Viel_WDM_Lya_2005,Viel_WDM_Lya_2008,Boyarsky_WDM_Lya_2009}.
More recently attention has turned to the large-scale forest's ability
to constrain the baryon acoustic oscillation peak, providing an
independent distance measurement for constraining dark energy
\cite{Slosar09BAO,BOSS_Slosar13,FontRibera_BOSS_QSO_Lya_X_BAO_2013}.
In addition to probing the power spectrum in these ways, the observed
$z<5$ forest constrains the thermal state of the intergalactic medium
\cite{Becker07,Bolton08}, allowing various interesting processes to be
studied (such as helium reionization \cite{Becker11HeIIreionization}).

When considering the forest after reionization, it is standard
practice \cite{Croft98,McDonald00,McDonald03} to model the IGM
ionization state in the presence of a uniform background of
ultraviolet (UV) photons.  Direct constraints on the Lyman-$\alpha$
cloud temperatures \cite{Bolton13IGMtemp} dictate that collisional
ionization is unimportant except in systems that are dense enough to
be substantially self-shielded from the radiation.

However the assumption that the UV background is uniform is known to
be incorrect, since the constituent photons are actually generated by
galaxies and quasars. One can distinguish two limits in which the
approximation fails. First, on small scales, quasars are rare;
depending on the fraction of photons they contribute (likely around
$50\%$ for $2<z<3$ \cite{Faucher-Giguere09_UVB,HM12}) they can add
significant shot noise on small scales.  Further fluctuations are
imprinted by intrinsic variability in the IGM opacity
\cite{Maselli05_LyA_transmissivity}.  This and related astrophysical
effects have been widely investigated elsewhere
\cite{Zuo92,Kollmeier03_LyA_LBG_connection,Meiskin04_LyA_radiation,Kollmeier06_LyA_galwinds,Viel_WDM_Lya_2008,Slosar09BAO,Mesinger09_LyA_fluctuating_reionisation,White10_LyA_sims,FontRibera_BOSS_QSO_Lya_X_2013}
with the conclusion that, if properly accounted for, the added noise
is not problematic for observational cosmology at $z<5$. Measurements
at higher redshift, during the epoch of reionization, will be affected
more strongly
\cite{Meiskin04_LyA_radiation,Mesinger09_LyA_fluctuating_reionisation}
as the UV undulation amplitude increases.

In this paper I will consider the post-reionization IGM and place more
emphasis on a second failure of the uniform-radiation assumption. This
appears only when source clustering is taken into account on scales
around the mean-free-path of an ionizing photon. By definition,
regions separated by greater distances cannot efficiently
exchange UV radiation. Ionization equilibrium will therefore depend on
the density of sources in the local region; the bias of the forest on
the largest scales will depend on the clustering of UV sources
\cite{Croft04_LyA_large_radiation,McDonald05,McQuinn11_LyA_fluctuations,White10_LyA_sims}.

This effect has received less attention to date, probably because the
relevant scale is seemingly extremely large (the mean-free-path is of
order $500\,\Mpc$ in comoving units \cite{Rudie13MFP} at $z=2.4$). In
fact, once redshifting and volume dilution are accounted for,
the transition scale is somewhat smaller (more
like $350\,\Mpc$ comoving; see Section \ref{sec:absorption-terms}). To
fully model such scales would require exceptionally large
radiative-transfer simulations, with box sizes exceeding a gigaparsec
to properly probe long-wavelength fluctuations.  

\updated{To achieve this, previous work has employed a combination of
  large dark-matter-only boxes and smaller hydrodynamic simulations
  \cite{Croft04_LyA_large_radiation} or semi-analytic prescriptions
  \cite{McDonald05}. In the former case the author reported a
  significant drop in large-scale flux power out to scales of $k^{-1}
  \sim 70\, h^{-1} \,\Mpc$ relative to the homogeneous-radiation
  control case. Even so, the result has not received widespread
  attention. This is likely because extending such state-of-the-art
  work consumes a great deal of computer time and, furthermore,
  appropriate empirical constraints for such large separations have
  seemed out of reach.}

The observational situation has now been changed radically by the BOSS
(Baryonic Oscillation Spectroscopic Survey) project
\cite{2013Dawson_BOSS_overview}. The team have released results
demonstrating the viability of measuring the correlation function of
Lyman alpha clouds on large scales
\cite{BOSS_Slosar11,2013_Busca_BOSS_Lya_BAO,BOSS_Slosar13}. The major
goal of BOSS is to measure the baryonic acoustic oscillation (BAO)
feature in the correlation function at $100\,h^{-1}\,\Mpc$
comoving. This is not so far off the reduced mean-free-path scale
discussed above and derived in Section~\ref{sec:absorption-terms}. It
is timely, therefore, to reconsider the impact of large-scale
fluctuations in the UV source density on the Lyman-$\alpha$ forest.
 
The remainder of this paper proceeds systematically from first
principles to a detailed linear-theory calculation of these
effects. This should be highly complementary to numerical studies, and
motivate further work in the area. I will ignore observational
questions such as the transformation from \hi to flux power spectrum,
redshift-space distortions, redshift evolution and flux calibration
biases -- since these require major computational machinery in
themselves \cite{2013_Busca_BOSS_Lya_BAO,BOSS_Slosar13} -- and focus
on the bias of the physical intergalactic \hi density at a single,
fixed redshift. The quantitative results will be presented for
$z=2.3$, around the mean redshift of observed Lyman-$\alpha$ clouds
\cite{2013_Busca_BOSS_Lya_BAO}.

The approximations that allow this calculation to be completed are
{\it (i)} that the spatial variations in the UV spectrum are less
important for \hi than the changes in intensity (a `monochromatic
approximation'); {\it (ii)} that the hydrogen can be split into a
diffuse intergalactic component in photoionization equilibrium and a
small population of self-shielded, collisionally-ionized clumps
(i.e. the highest-column-density Lyman limit systems
\cite{McQuinn_11_LyLimit}); {\it (iii)} that non-linear corrections
(including quasar duty cycles) can be ignored on sufficiently large
scales
\cite{Slosar09BAO,Mesinger09_LyA_fluctuating_reionisation,White10_LyA_sims},
although I will include shot noise from the rarity of sources; {\it
  (iv)} that sources averaged on large scales radiate
isotropically. These seem reasonable to obtain a good estimate of the
effects but in future they should be checked against numerical
simulations \updated{and more complicated analytic treatments that allow
  for departure from equilibrium \cite{Zhang07_reion}}.
After circulating a draft of this work, I was made aware of an
independent study by Gontcho A Gontcho, Miralda-Escud\`e and Busca (in
prep); at present it seems these authors reach many similar
conclusions using a different calculation framework. This is
encouraging, and it will be helpful to compare our approaches in
due course.

Section \ref{sec:nonrel-radiative-transfer} develops the
inhomogeneous, monochromatic radiative transfer equations; Section
\ref{sec:linearisation} discusses the application of these equations
to the large-scale, linear behaviour of intergalactic \hi. Section
\ref{sec:results} presents the main results, showing how various
parameters change the distinctive imprint of radiative transfer on the
intergalactic neutral hydrogen. Further discussion is given in Section
\ref{sec:discussion}, especially in relation to observations of the
Lyman-$\alpha$ forest.  Two subsidiary issues are considered
in appendices. In Appendix \ref{sec:adding-gravity}, I re-derive all
equations using general relativity, so including peculiar velocities
and inhomogeneous gravitational redshifting and elucidating the gauge-dependence of the
results (all of which considerations turn out to impact only on scales larger than
those of interest here). Appendix \ref{sec:estimate-b_0} discusses the
calibration of a particular parameter (the intergalactic \hi bias in
the absence of radiation transfer) from analytic arguments and numerical simulations.

There are a few notational matters worth settling before starting the
calculation.  It is helpful to be able to decompose any quantity
$X$ into its spatial mean value $X_0$ and fractional perturbations
$\delta_X$ defined by
\begin{align}
\delta_X = \frac{X-X_0}{X_0}\textrm{.}\label{eq:define-delta}
\end{align}
Later I will mainly
deal with the Fourier transform $\tdelta_X$ of these fractional variations; any
quantity can be re-written
\begin{equation}
\delta_X(\vec{x}) = \frac{1}{(2\pi)^{3/2}}\int \dd^3 k\, e^{i \vec{k} \cdot \vec{x}}
\tdelta_X(\vec{k})\textrm{,} \label{eq:FT-delta}
\end{equation}
where $\vec{k}$ is the comoving wavevector. Finally, the
power spectrum $P_X(k)$ is defined by
\begin{equation}
\langle \tdelta_X(\vec{k}')^* \tdelta_X(\vec{k}) \rangle = P_X(k) \delta(\vec{k}-\vec{k}')\label{eq:define-Pk}
\end{equation}
where angle brackets denote an ensemble average and, by an unfortunate
quirk of conventional notation, the $\delta$ on the right hand side
represents the Dirac delta function. It follows from these definitions
that the power spectrum for any quantity has units of a comoving
volume.  The expression above assumes statistical isotropy so that
$P_X$ is a function of $k=|\vec{k}|$ alone.

Numerical results will be derived assuming a fiducial Planck
temperature-only \cite{Planck13_CosmoPar} cosmology
$(h,\Omega_{M0},\Omega_{\Lambda0})=(0.6711,0.3175,0.6825)$, where
$\Omega_{M0}$ and $\Omega_{\Lambda0}$ are the present day matter and
cosmological constant densities relative to critical and
$h=H_0/(100\,\km\,\mathrm{s}^{-1}\,\Mpc^{-1})$ is the dimensionless
Hubble parameter today. The main role of these quantities will be to
fix the Hubble expansion rate at $z=2.3$; any uncertainties are
easily small enough to be ignored for the present study.

\section{Radiative Transfer}\label{sec:nonrel-radiative-transfer}

In this Section, I will derive a monochromatic approximation to the
radiative transfer equation; this involves systematically integrating
over frequency dependence. Because the scales of interest remain
strongly sub-horizon, relativistic corrections will be sub-dominant and
are excluded. For the interested reader, they are reintroduced in
Appendix \ref{sec:adding-gravity} which shows explicitly that they
constitute a small correction.

To start, let $f(\vec{x}, \vec{n}, \nu)$ denote the physical number
density of photons at comoving position $\vec{x}$ traveling in
direction $\vec{n}$ with frequency $\nu$. In the absence of
collisional effects, the total number of photons is conserved. However
the Lagrangian phase volume that those photons occupy changes over
time: the spatial volume increases as $a^3$ while the frequency
interval decreases as $a$, giving an overall expansion
rate\footnote{Some works, e.g. Refs
  \protect\cite{1996ApJ...461...20H,HM12}, choose to use the energy
  density per unit volume, which leads to an $a^3$ volume factor and
  accordingly a few cosmetic differences.} of $a^2$. Overall, this
implies the following Boltzmann equation:
\begin{equation}
  \frac{\partial f}{\partial t} + \frac{c}{a}
  (\vec{n} \cdot
  \nabla) f + \frac{\partial f}{\partial \nu} \frac{\dd
    \nu}{\dd t} + 2H f = C[f]\textrm{,}\label{eq:Boltzmann}
\end{equation}
where $c$ is the speed of light, $a$ is the universe scalefactor and
$H=\dot{a}/a$ is the usual Hubble expansion rate. $C[f]$ contains the
collisional terms ({\it i.e.} those that alter the photon number) and
will be expanded in a moment. In order, the terms on the
left-hand-side denote the Eulerian rate of change of photon density;
the free-streaming of photons; the redshifting of the photons; and the
volume dilution discussed above.  The gradient operator $\nabla$ is
taken with respect to the comoving position $\vec{x}$ throughout this
work. The term $\partial f/\partial t$ will now be set to zero,
meaning I am approximating the radiation and ionization to be in
equilibrium as noted in the Introduction. At the background level,
this is a good approximation at $z=2.3$ -- the evolution of the
photoionization rate $\Gamma_0$ is slow, $\dd \ln \Gamma_0 / \dd
\ln a \approx -0.04$ from the tabulations of Ref. \cite{HM12} -- but
the implications of time-dependence for perturbations should certainly
be explored further in future work.

To formulate the collisional term, consider first the emission of
radiation. There are two distinct relevant aspects: first, galaxies
and quasars generate energy from stars and black holes; second, the
intergalactic \hi regenerates a fraction of photons it previously
absorbed when the electron and proton recombine. I will treat these
two terms separately in what follows.

Now consider absorption processes. A large effect will come from
the IGM, corresponding to the low-column-density Lyman-$\alpha$
forest. The density of the neutral hydrogen $\nHI(\vec{x})$ in this
phase will be a key quantity. However, some portion (to be quantified
later) of absorption comes from small, dense clumps which are strongly
self-shielded against the UV radiation that is being modeled. At
least three characteristics distinguish the clumped phase: first, the
density of \hi is determined by collisional ionization and hence
essentially unaffected by variations in the radiation. Second, the
majority of recombination radiation produced is re-absorbed internally
within a clump. Third, the amount of radiation absorbed by such a
population does not scale with the mass of \hi in the population, but
rather with the geometrical size and number density of the objects.
For all three reasons, this population requires separate treatment.

Following the above discussion, the emission and absorption of photons
can be expressed by
\begin{align}
C_{\nu}\left[f\right]  = j_{\nu}(\vec{x}) & + \nHI(\vec{x})
\left(\frac{\Gamma(\vec{x})}{4 \pi} \frec(\nu,T) - c \sigHI\left(\nu
  \right) \, f  \right) \nonumber \\
 & - c \kappaclump(\vec{x},\nu)f\textrm{,}\label{eq:collision-terms}
\end{align}
where:
\begin{itemize}
\item $j_{\nu}(\vec{x},\vec{n})$ is the emissivity per unit physical volume per
  frequency interval from sources other than the IGM itself;
\item $\nHI(\vec{x})$ is the number of ground-state hydrogen atoms per
  unit physical volume in the IGM (excluding the shielded clumps);
\item $\kappaclump(\vec{x},\nu)$ is the opacity from
  collisionally-ionized clumps;
\item $\frec(\nu,T)$ is the IGM recombination spectrum, which depends on the
  temperature $T$ of the free electrons;
\item $\Gamma(\vec{x})$ is the rate of ionization per \hi atom (and
  therefore also the recombination rate, assuming photoionization
  equilibrium); and
\item $\sigHI(\nu)$ is the cross-section of a \hi atom to ionization
  by a frequency $\nu$ photon.
\end{itemize}
In principle $j_{\nu}$ is a function of angle $\vec{n}$ as well as of
position $\vec{x}$ but, in accordance with approximation {\it (iv)}
above, the $\vec{n}$ dependence is now to be dropped (meaning that
sources averaged over large scales radiate isotropically).  I will
also assume throughout that only \hi can absorb photons in the
frequency range of interest.

The cross-section $\sigHI$ is sharply peaked at the Lyman limit
($\nuLL \approx 3 \times 10^{15}\,\mathrm{Hz}$), which allows for a
monochromatic approach. The key quantity will be an effective number
density of Lyman limit photons, $\fLL$, defined by
\begin{equation}
\fLL(\vec{x},\vec{n}) = \int f(\vec{x},\vec{n},\nu) \sigHI(\nu)\,\dd\nu\textrm{.}
\end{equation}
If desired, one can divide through by a fixed cross-section
(e.g. $\sigHI(\nuLL)$) to ``correct'' the units of $\fLL$, leading to
cosmetic differences.  Either way, $\sigHI$ defines a single
particular broad-band filter that we choose to focus on; the whole
framework could be formulated in terms of another band if
desired. This particular choice of filter is uniquely motivated
because the ionization rate per \hi atom -- a critical quantity of
interest -- is given exactly by integrating $\fLL$ over all angles:
\begin{equation}
\Gamma(\vec{x}) = c \iint \fLL(\vec{x},\vec{n}) \, \dd^2 n\textrm{.}\label{eq:ionization-rate}
\end{equation}

To obtain the Boltzmann equation for $\fLL$, multiply equation
\eqref{eq:Boltzmann} by $\sigHI$ and integrate with respect to $\nu$,
giving 
\begin{equation}
  a^{-1} (\vec{n} \cdot \nabla) \fLL +
  \kappat \fLL = c^{-1}\left(j + \sigHIbar \nHI \frac{\Gamma}{4 \pi} \betarec(T) \right) \textrm{,}\label{eq:fLL-boltzmann}
\end{equation}
where $\kappat$ is an effective opacity, $j$ is an effective source
emissivity and $\betarec(T)$ is a dimensionless, temperature-dependent
fraction of recombination radiation which lies in our Lyman limit
waveband. The formal definitions of these terms arise directly from the
frequency integration, and will be given and discussed below in turn.

\subsection{Absorption}\label{sec:absorption-terms}

First consider the effective opacity $\kappat$ which has been composed
from separate diffuse IGM opacity, clump opacity, redshifting and
volume dilution contributions:
\begin{equation}
\kappat = \sigHIbar \nHI + \kappaclumpbar + \alpha_z \frac{H}{c} + 3\frac{H}{c}\textrm{.}
\end{equation}
I have written the volume term as $3H/c$ to directly associate it with
comoving volume dilution; $\alpha_z$, a dimensionless number to be
defined below, will contain a compensating term to return the $2H/c$
of the original formulation \eqref{eq:Boltzmann}.  The quantities
$\sigHIbar$ and $\alpha_z$ are dependent on the spectrum, but not on
the normalization of the spectrum; the monochromatic approach
therefore assumes them independent of position.  Their values can be
estimated by using tabulated mean UV background estimates
\cite{HM12} at $z=2.3$:
\begin{align}
\sigHIbar &= \frac{1}{\fLL} \int \sigHI^2 f \, \dd \nu \approx 3.87 \times
10^{-18}\,\cm^{-2}\textrm{;}\\
\alpha_z &= -\frac{1}{\fLL} \int \sigHI \frac{\partial f}{\partial \ln
  \nu} \, \dd \nu - 1 \approx 1.57 \textrm{.}\label{eq:alphaz}
\end{align}
Meanwhile I have defined the monochromatic clump opacity
\begin{equation}
\kappaclumpbar(\vec{x}) = \int \kappaclump(\vec{x},\nu) \, \sigHI(\nu)
\, \dd \nu\textrm{.}
\end{equation}
It will be convenient later to write the fraction of effective opacity
from the respective terms as
\begin{equation}
\betaHI = \frac{\sigHIbar \nHI}{\kappat}\textrm{; }
\betaclump = \frac{\kappaclumpbar}{\kappat}\textrm{; }
\betaz = \frac{\alpha_z H}{c \kappat}\textrm{; }
\betaV = \frac{3H}{c \kappat}\textrm{.}\label{eq:betas}
\end{equation}
By definition these obey $\betaHI+\betaclump+\betaz+\betaV=1$. We can
estimate their values by referring to the observational constraints on
Lyman limit opacity; for instance Ref. \cite{Rudie13MFP} quote a
mean-free-path of $\kappaHI^{-1} \equiv (\sigHIbar
\nHI+\kappaclumpbar)^{-1} \approx 150\,\Mpc$ in physical units at
$z\approx 2.4$. Their value takes into account the intergalactic
medium absorption alone (it excludes volume and redshifting effects,
as well as circumgalactic absorption immediately around the emitting
object).  Correcting to $z=2.3$ using \cite{Rudie13MFP}
$\lambda_{\mathrm{MFP}} \propto (1+z)^{-4.5}$ and converting to
comoving units gives a helpful reference value:
\begin{equation}
a^{-1}\kappaHI^{-1} \approx 570\,\Mpc\textrm{ comoving at }z=2.3\textrm{.}
\end{equation}
There are uncertainties in the analysis of the observational data
\cite{Prochaska13_MFP} which could imply that the correct
mean-free-path is somewhat longer than this value; results for
different $\kappaHI$ will be investigated at the end of
the work.

With the Planck cosmology defined in the introduction one has
$H(z=2.3)/c \approx (1280\,\Mpc)^{-1}$ and therefore, taking the
reference value of $\kappaHI$ above,
\begin{equation}
\betaHI + \betaclump = 0.62\textrm{; } \beta_z=0.13 \textrm{; } \betaV=0.25\textrm{.}
\end{equation}
One immediate implication of these calculations is that, compared
against physical opacity, redshifting and volume dilution are
sub-dominant but important factors in lowering the cosmological density
of Lyman-limit photons. This implies that the relevant scale at which
scale-dependent effects are centered is smaller than the quoted
\hi-only mean-free-path; we now have
\begin{equation}
a^{-1}\kappat^{-1} \approx 350\,\Mpc\textrm{ comoving at }z=2.3\textrm{.}
\end{equation}
This is closer to the range measurable by BOSS. In fact when solving
the equations in detail below, this characteristic path-length will turn out
to be sufficiently short that radiation transfer can have a significant impact on
the forest at the BAO scale.

Finally, one needs to decide how to assign opacity between the
intergalactic \hi and clumps. Sadly there is no way to do this
unambiguously so I will further parameterize:
\begin{equation}
\pclump = \frac{\betaclump}{\betaclump + \betaHI}\textrm{.}
\end{equation}
Ref. \cite{Rudie13MFP} details the fraction of opacity from systems of
differing column density, allowing an estimate of $\pclump$.  For a
parcel of gas to count as clumped, the definition made above equation
\eqref{eq:collision-terms} requires it to be in collisional ionization
equilibrium. The boundary will therefore be somewhat higher than the
traditional Lyman limit system threshold because reaching the
collisionally-ionized state requires a reduction in photoionization
rate by a substantial fraction throughout the cloud. 

The details depend strongly on the temperature, density and geometry
of the system itself; here I will make a very rough order-of-magnitude
estimate for a cut-off point. From the radiative-transfer simulations
in Ref. \cite{PontzenDLA}, I determined that a typical Lyman-limit
system has an electron density around $10^{-2}\,\cm^{-3}$ and a
temperature $T\approx 2.4 \times 10^{4} \mathrm{K}$.  This gives a collisional ionization rate of
approximately \cite{Black81clouds} $1.2 \times 10^{-13} \mathrm{s}^{-1}$,
compared to the photoionization rate \cite{HM12} of around $1.0 \times
10^{-12} \mathrm{s}^{-1}$. One therefore needs to suppress the
intergalactic flux by a factor of around ten to reach collisional ionization.

Then, making a simple uniform-density 1D model of a clump irradiated
by the intergalactic flux, the mean flux inside as a function of total
column density $N$ is given by $\Gamma_0 (1-e^{-\sigHIbar
  N})/(\sigHIbar N)$. Note therefore that, although the central
photoionization rate falls exponentially with $N$, the mean falls only
approximately linearly with $N$. Accordingly for the mean rate to drop
to $\Gamma_0/10$ requires $N\approx 10 \sigHIbar^{-1} \approx (2.6
\times 10^{18})\,\cm^{2}$.  The cumulative effect of column densities
greater than these limits constitute only around 10\% of the IGM
opacity (see Ref.  \cite{Rudie13MFP}, Figure 10). For that reason I
will adopt $\pclump = 0.10$, showing the effect of varying the value
at the end of the paper.

\subsection{Recombination emission}

Now let us turn attention to the recombination radiation term. This
arises automatically from the integration discussed above equation
\eqref{eq:fLL-boltzmann}, with the definition
\begin{equation}
\betarec(T) = \frac{1}{\sigHIbar} \int_0^{\infty}
\sigHI(\nu) \frec(\nu,T) \, \dd \nu\textrm{,}\label{eq:beta-rec}
\end{equation}
which provides a dimensionless measure of the
amount of recombination radiation that ends up in the monochromatic
waveband under consideration. Over the frequencies of interest, $\frec$
can be approximated as an offset Maxwell-Boltzmann distribution corresponding
to the electron temperature $T$, scaled by the fraction of
recombinations that occur directly to the ground state:
\begin{equation}
\frec(\nu,T) \approx 2h f_{\infty\to 1} \sqrt{\frac{h (\nu-\nuLL) }{\pi (kT)^3}}
e^{-h(\nu-\nuLL)/kT}\textrm{,}\label{eq:Maxwell-Boltzmann}
\end{equation}
where $f_{\infty\to 1}\approx 0.40$ is the fraction of recombinations
direct to the ground state \cite{loeb2013firstgals}, 
$h$ is Planck's constant and $k$ is Boltzmann's constant. At $z\approx
2.4$ (close enough to our fiducial redshift), a typical forest
temperature is \cite{Becker11HeIIreionization,Bolton13IGMtemp} $T=2.5
\times 10^4\,\mathrm{K}$; evaluating equation \eqref{eq:beta-rec} then
gives $\betarec = 0.39$. In principle, we could keep track of how
variations in the mean temperature correlate with variations in the
mean density; note, however, that when considering the averaged
effects on linear scales this may not be the same as the equation of
state measured for individual clouds \cite{Bolton13IGMtemp}. Worse,
large scale spatial temperature correlations could well be generated
by unmodeled, non-equilibrium phenomena such as helium reionization
\cite{Lai05_LyA_temp_flucs,Bolton06_UV_shape_fluctuations,Furlanetto09_He_fluc,Becker11HeIIreionization}. Luckily,
the final effect of these on the photoionization equilibrium will be
sub-dominant because $\betarec$ changes quite slowly with temperature
($\dd \betarec / \dd \ln T = -0.15$). \updated{The impact of thermal
  fluctuations on the recombination spectral shape is thus small
  compared to their effect on the recombination rate (which scales
  approximately as $T^{-0.7}$ in the intergalactic regime). Even in
  the latter case, within our monochromatic approximation the
  temperature variations do not depend strongly on the local ionizing
  field strength \cite{HuiGnedin97_IGM_EoS}.  Incorporating
  multi-wavelength, time-dependent radiative transfer could introduce
  qualitatively important corrections to the temperature field and
  should be prioritized in future work (see Section
  \ref{sec:discussion}). }

\subsection{Other sources}\label{sec:emissivity}

There is one remaining term in equation~\eqref{eq:fLL-boltzmann} that
as-yet has not been discussed: $j(\vec{x})$. Recall
that~\eqref{eq:fLL-boltzmann} is obtained by
integrating~\eqref{eq:Boltzmann} over frequency; accordingly
$j(\vec{x})$ is defined by
\begin{equation}
j(\vec{x}) = \int j_{\nu}(\vec{x}) \, \sigHI(\nu) \, \dd \nu\textrm{,}
\end{equation}
and specifically excludes the recombination emission which was treated
separately above.  Looking ahead in the calculation, we will need to
understand the statistical properties of the $j(\vec{x})$ field.  It
is widely believed that, at $z\approx 2.3$, quasars and galaxies both
contribute significantly to the UV emission \cite{HM12}. For both
populations, systematic fluctuations $\delta_j$ are thought to be
proportional to a constant (the `bias', $b_j$) times the matter
density fluctuations $\delta_{\rho}$ \cite{2011PhRvD..84f3505B} when
averaged over suitably large scales. However we will also need to
consider the shot noise: because quasars are rare, even a uniform
distribution would have a significant Poisson fluctuation in density
from place to place. In the limit of large scales (and therefore large
numbers), these Poisson fluctuations can be modeled as an additive
Gaussian noise, giving the total large-scale emissivity variations:
\begin{equation}
\delta_j(\vec{x}) = b_j\, \delta_{\rho}(\vec{x}) +
\delta_{\mathrm{SN}}(\vec{x}) \textrm{,}\label{eq:j-decomposition}
\end{equation}
where, in accordance with definition \eqref{eq:define-delta},
$\delta_{\rho}(\vec{x})$ is the fractional matter overdensity
determined by the cosmology and $\delta_{\mathrm{SN}}(\vec{x})$ is the
uncorrelated shot-noise Gaussian random field. If multiple source populations
contribute to the emissivity, they add linearly from which it follows
that
\begin{equation}
\delta_j(\vec{x}) = \sum_i \frac{j_{0,i}}{j_0} \left( b_{j,i}\,
  \delta_{\rho}(\vec{x}) + \delta_{\mathrm{SN,i}}(\vec{x})\right)\textrm{,}\label{eq:j-decomposition-multiple}
\end{equation}
where the sum extends over the different source categories $i$ and
the total mean emissivity is $j_0=\sum_i j_{0,i}$. Comparing equations
\eqref{eq:j-decomposition} and \eqref{eq:j-decomposition-multiple}
shows that the net large-scale effect is the same as that of a single
population with suitably averaged parameters as I will discuss below.

Consider first the bias $b_j$, which is established by estimating the
correlation strength of the emitting objects. Quasars are found to be
strongly biased ($b_q \approx 4$) with respect to the matter density
field \cite{Croom2dF_QSObias05,White_QSO_clustering}. Galaxies are
substantially less strongly correlated, and therefore less biased; Ref.
\cite{Adelberger05_galaxy_correlation} quotes a correlation length of
$r_0=4.3\,h^{-1}\,\Mpc$ for a sample of bright
($23.5<\mathcal{R}<25.5$) galaxies at $z\approx 2.2$. This translates
into a bias of $b_g \approx 2.4$ with the Planck cosmology described
above, assuming the underlying matter fluctuations to be normalized
\cite{Planck13_CosmoPar} to $\sigma_8 = 0.834$ at $z=0$.

To add complication, these biases are measured for bright objects;
especially in the case of galaxies, a significant fraction of photons
are emitted from a large population of individually under-luminous
objects \cite{Reddy09_UV_LF}. To understand how the bias scales with
luminosity one can assume it arises from the underlying dark matter
halo. In that case the bias implies a halo mass
\cite{ColeKaiser_Bias_89}; for instance, with $b_g=2.4$ we obtain a
characteristic mass\footnote{These results have been calculated using
  the prescriptions of Ref. \cite{ColeKaiser_Bias_89} as implemented by
  Ref. \cite{2013ascl.soft05002P}.} of $M = 4\times 10^{11}\,\Msol$.  Suppose
we wish to consider galaxies a factor of $10$ fainter than the sample
of Ref. \cite{Adelberger05_galaxy_correlation}; then a number of
arguments point to the dark matter halos being approximately a factor
of $\sqrt{10}$ less massive \cite{Moster13,PontzenDLA}. This can be
translated back into a bias of $1.9$. Similarly, dropping another
factor of $10$ in luminosity yields halos with bias $1.5$.

So the appropriate `source bias' is sensitive to details of the
underlying population generating the UV photons. For a combination of
different sources, equation \eqref{eq:j-decomposition-multiple} shows
that the net bias is exactly the average of the two individual biases,
weighted by the emissivity of the populations. As a default value in
this work, I will assume an average source bias of $b_j=3$,
representing the average between highly biased quasars and a range of
galaxy luminosities. (Recall that the value does not need to be
further reduced for recombination emission, since that is included
elsewhere in the calculation.) Reflecting the uncertainty, I
will also show results for a range of $b_j$ from $1.5$ to $4.0$. A
great attraction of future measurements of the effects in this paper
is that they should in principle constrain $b_j$ and so shed light on
the origin of UV photons.

Now consider the shot-noise term $\delta_{\mathrm{SN}}$ for a single
population. This represents random variations in the density of
sources. For a mean density of $\bar{n}$, the number of sources in a
volume $V$ is given by
\begin{equation}
N(V) = \bar{n}\left(V + \frac{1}{(2 \pi)^{3/2}} \int_V \dd^3 x \int
  \dd^3 k \, e^{i
    \vec{k}\cdot\vec{x}} \tdeltaSN(\vec{k})\right)\textrm{.}
\end{equation}
Using this expression to demand that $\langle N^2 \rangle - \langle N
\rangle ^2 = N$ for any volume $V$ (the Gaussian, large-$N$ limit of
Poisson noise) dictates the power spectrum
\begin{equation}
P_{\mathrm{SN}}(k) = \bar{n}^{-1}\textrm{,}
\end{equation}
independent of scale, where $P_{\mathrm{SN}}(k)$ is defined by
equation \eqref{eq:define-Pk} and correctly has units of comoving
volume as explained earlier. The shot-noise is modeled as stationary;
in fact quasars likely have a finite duty cycle, causing the
realization of the shot-noise to change over time. The effect of this
cannot be analyzed rigorously with the time-stationary approach I have
adopted, but it could plausibly change the impact of noise on large
scales. It should therefore be investigated in future work.

Just as for the bias $b_j$, choosing an appropriate value of $\bar{n}$
is tricky. \updated{One can start by parameterizing the quasar
  luminosity function $\Phi(L)$ using a double power-law fit
  \cite[e.g][]{2007ApJ...654..731H,Ross13_SDSS9_QLF}; following the
  consequences of equation \eqref{eq:j-decomposition} for a series of
  infinitesimal bins in luminosity, assuming an independent
  shot-noise realization for each bin, one is led to an $L^2$-weighted
  \cite{Mesinger09_LyA_fluctuating_reionisation,McQuinn11_LyA_fluctuations}
  effective quasar number density defined by
\begin{equation}
\bar{n}_q = \frac{\left(\int \Phi(L)\, L\, \dd L\right)^2}{\int \Phi(L)\, L^2\, \dd L}\textrm{.}\label{eq:nq-from-PhiL}
\end{equation}
Since the intrinsic luminosity function in the ionizing radiation is
unknown (being completely obscured by Lyman limit absorption) I assume
that the ionizing radiation of a given quasar scales linearly with its
bolometric luminosity. Evaluating equation \eqref{eq:nq-from-PhiL}
using estimates for the bolometric population parameters
\cite{2007ApJ...654..731H} then gives $\bar{n}_q$ between $1.5 \times
10^{-6}$ and $10^{-5}\,\Mpc^{-3}$ comoving over the parameter range
quoted by Ref \cite{2007ApJ...654..731H}.}

As with $b_j$, the $\deltaSN$ appearing in equation
\eqref{eq:j-decomposition} can be seen from equation
\eqref{eq:j-decomposition-multiple} to be a photon-weighted average of
the $\deltaSN$ appropriate to the two populations. The density of
galaxies is so much higher that one can essentially ignore their
shot-noise contribution compared to that of the quasars.  Tracing this
through, assuming again a 50\% contribution from both populations, one
has to multiply $\bar{n}$ by 4 to account for the galaxy part of the
emission (since $\deltaSN$ scales with $\bar{n}^{-1/2}$).  This yields
the approximate upper limit $\bar{n} \approx 4 \times
10^{-5}\,\Mpc^{-3} \approx 10^{-4}\,h^3\,\Mpc^{-3}$, with a lower
limit of $2\times 10^{-5}\,h^3\,\Mpc^{-3}$. To be clear, the $\bar{n}$
derived in this way is not the density of any particular population --
it is a weighted average which accounts for the very different
densities of two populations. 

These estimates neglect any effects of time-variability and anisotropy
which will introduce qualitative corrections and possibly lead to an
increase in the effective $\bar{n}$ by pushing observed variation to
smaller scales.  I will therefore adopt the upper end of the naive
uncertainty for the present. Results will later be shown for a full
range of possible $\bar{n}$.  Many of the same physical considerations
bear on the value of both $\bar{n}$ and $b_j$, but I will consider
them to be separate parameters for the sake of clarity.

The total power spectrum for the sources $P_j(k)$ is just
\begin{equation}
P_j(k) = b_j^2 P_{\rho}(k) + \bar{n}^{-1}\textrm{,}\label{eq:Pj}
\end{equation}
which follows because, by assumption, $\langle \tdelta_{\rho}(\vec{k})
\tdeltaSN(\vec{k}') \rangle = 0$. A similar expression was given by
Ref. \cite{McQuinn11_LyA_fluctuations}. However, despite appearing on
an equal footing in equation \eqref{eq:Pj}, the shot-noise and
correlated components behave differently in terms of their effect on
the \hi \cite{White10_LyA_sims}, as we will see below.

\section{Linearization}\label{sec:linearisation}

The preceding section concluded by discussing the behaviour of sources
averaged on large scales in terms of spatial perturbations $\delta$
defined by \eqref{eq:define-delta}. The plan now is to rewrite the
radiative transfer equations in terms of $\delta$'s. Ignoring all
terms of order $\delta$ yields the homogeneous or ``zero-order''
approximation; the Boltzmann equation \eqref{eq:fLL-boltzmann} becomes
(integrating over all angles without loss of information)
\begin{equation}
\kappatb (1-\betaHI \betarec) \Gammab = 4\pi j_0 \textrm{,}\label{eq:background-equilibrium}
\end{equation}
which expresses the equilibrium condition that the overall rate of
photon production is balanced by the effective absorption from
redshifting, dilution and ionization. 

Expanding equation~\eqref{eq:fLL-boltzmann} to linear order and
simplifying using the background solution~\eqref{eq:background-equilibrium},
one obtains an expression for $\tdelta_{\fLL}$:
\begin{equation}
\hspace{0.7cm} 
\tdelta_{\fLL} = \frac{(1-\betaHI\, \betarec)\,\tdelta_{j} + \betaHI\, \betarec \,\left[ \tdelta_{\nHI} +   \tdelta_{\Gamma}\right] - \tdelta_{\kappat}}{i(a\kappatb)^{-1}\, (\vec{n} \cdot \vec{k}) + 1 }\textrm{,}\label{eq:sln-with-angular-deps}
\end{equation}
where I have suppressed functional $\vec{k}$ and $\vec{n}$
dependencies for brevity. Excepting small gravitational effects
(Appendix \ref{sec:adding-gravity}), the effective opacity
fluctuations $\delta_{\kappat}$ are linked solely to variations in the
diffuse and clumped neutral hydrogen:
\begin{equation}
\delta_{\kappat} = \betaHI \delta_{\nHI} + \betaclump \delta_{\kappaclumpbar}\textrm{.}\label{eq:delta-kappat-no-grav}
\end{equation}

\begin{figure}
\includegraphics[width=0.5\textwidth]{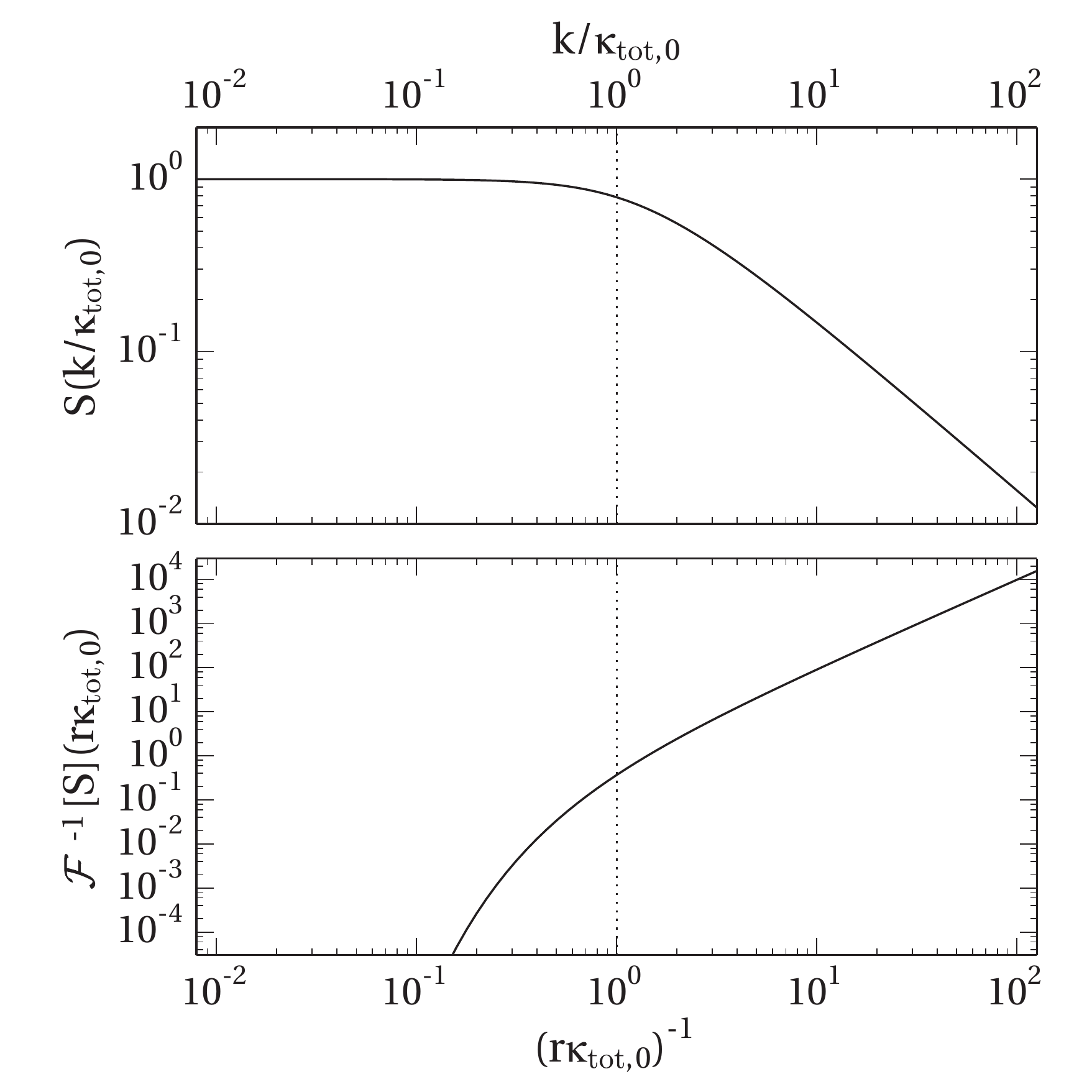}
\caption{In the linear approximation, the radiative transfer consists
  of convolving an effective source function (including emission,
  absorption and re-radiation terms) with a kernel {\it S}, equation
  \eqref{eq:radiation-equation}. The kernel is shown here in Fourier
  space (upper panel) as a function of wavenumber divided by
  $\kappatb$. On large scales (toward the left) $S(k)$ is $\approx 1$,
  meaning fluctuations in the effective source function are tracked by
  fluctuations in the number density of ionizing photons.  On small
  scales (toward the right), $S(k)$ decays towards zero; fluctuations
  in the photon density are suppressed and the uniform UV
  approximation will apply. The lower panel shows the same kernel
  transformed into real space; the horizontal axis is an inverse
  distance, so the two panels read in the same direction. }\label{fig:Sk}
\end{figure}

Integrating out the remaining angular dependence in equation
\eqref{eq:sln-with-angular-deps}, noting that $\int \dd^2 n\,
\tdelta_{\fLL} = 4 \pi \tdelta_{\Gamma}$, one obtains an implicit
equation for~$\tdelta_{\Gamma}$:
\begin{align}
\tdelta_{\Gamma}(\vec{k}) &= \left[(1-\betaHI\, \betarec)\, \tdelta_j  -
  \betaHI\,(1-\betarec)\, \tdelta_{\nHI} \right. \nonumber \\
&\left. \hspace{2.0cm} - \betaclump\, \tdelta_{\kappaclumpbar} + \betaHI\, \betarec\, \tdelta_{\Gamma}\right] S(k)\textrm{,} \nonumber \\
S(k) &= \frac{a\kappatb}{k} \arctan \frac{k}{a\kappatb}\textrm{,}\label{eq:radiation-equation}
\end{align}
showing the characteristic scale-dependence arising in the radiation
field (Figure \ref{fig:Sk}, upper panel). Performing an inverse
Fourier transform on the kernel $S(k)$ returns the radial function
$e^{-\kappatb r}/r^2$ (Figure \ref{fig:Sk}, lower panel).  The
systematic approach has therefore recovered something like the
heuristic equations of Refs
\cite{Mesinger09_LyA_fluctuating_reionisation,McQuinn11_LyA_fluctuations},
where sources are convolved with a similar kernel.  However, those
works do not take into account the shortened effective mean-free-path
from redshifting and volume-dilution contributions (they use $\kappaHI$ where
$\kappatb$ is more appropriate). Moreover the appearance of
$\tdelta_{\nHI}$ (from inhomogeneous absorption) and $\tdelta_{\Gamma}$
(from recombination radiation) on the right-hand-side of equation
\eqref{eq:radiation-equation} means that the convolution
kernel is modified from this simple form once absorption fluctuations, as well as
emission fluctuations, are included.

\begin{table}
\begin{tabular}{llr}
  \hline 
  & {\it Spectrum-dependent coefficients} \\
  $\alpha_z$ & Coefficient for background redshifting
  \eqref{eq:alphaz} & $1.57$ \\
  $\betarec$ & Fraction of \hi recombinations to LL photons \eqref{eq:beta-rec}  & $0.39$
  \\
  \hline
  & {\it Estimated origin of effective opacity} $\kappatb$, eq. \eqref{eq:betas}  \\
  $\betaHI$ & Fraction from \hi in photoionization equilibrium &
  $0.56$ \\
  $\betaclump$ & Fraction from
  collisional-equilibrium clumps & $0.06$ \\
  $\beta_z$ & Fraction from redshifting &
  $0.13$ \\
  $\beta_V$ & Fraction from dilution & $0.25$ \\
  \hline
  & {\it Input biases relative to the linear overdensity field } \\
  $\bu$ & Bias of \hi in homogeneous radiation limit & $1.5$ \\
  $b_j$ & Bias of photon source objects & $3$ \\
  $\bjeff$ & Effective bias of sources including recombination
  \eqref{eq:bjeff-def} & $2.6$ \\ 
  \hline
\end{tabular}
\caption{Dimensionless quantities used in this work, with a brief
  a  explanation and the default value calculated or estimated at $z=2.3$.}
\end{table}

On large scales $S(k \to 0)$ is $\approx 1$, meaning fluctuations in
the effective source function are tracked by fluctuations in the
number density of ionizing photons. This agrees with the intuitive
picture outlined earlier, in which regions separated by more than the
mean-free-path must arrive at independent ionization equilibria. On
small scales $S(k \to \infty)$ decays towards zero; fluctuations in
the photon density are suppressed and the uniform UV approximation is
recovered. (On sufficiently small scales one will, however, have
enhanced non-linear shot noise; as discussed in the introduction I
will consider only the linear regime in the present work.)

We are now in a position to understand the mean ionization
state. Recalling that the effects of shielded clumps have already been dealt with --
the $\nHI$ field refers specifically to the intergalactic 
\hi alone -- we can write
\begin{equation}
\delta_{\nHI} = \deltau - \delta_{\Gamma}\textrm{,}\label{eq:gamma-to-nHI}
\end{equation}
\updated{where $\deltau$ describes the \hi field in the case of a completely
uniform ionizing background; the given relationship is a consequence
of linearizing the photo-ionization equilibrium equation $\nHI \propto
1/\Gamma$. In the absence of any radiative fluctuations, by definition
$\delta_{\nHI} = \deltau$.} Combining equation~\eqref{eq:radiation-equation}
and~\eqref{eq:gamma-to-nHI} gives a solution for $\tdelta_{\nHI}$ in
terms of $\tdelta_j$, $\tdeltau$ and $\tdelta_{\kappaclumpbar}$:
\begin{equation}
\tdelta_{\nHI} = \frac{ \tdeltau - 
  \left[(1- \betaHI \betarec)\, \tdelta_j - \betaclump
    \tdelta_{\kappaclumpbar} + \betaHI\, \betarec\, \tdeltau\right] S(k)}{1 - \betaHI\, S(k) }\textrm{.}\label{eq:delta-nHI-main}
\end{equation}
Now assume that $\tdeltau$ (the \hi density fluctuations in a
completely uniform UV field), $\tdelta_j$ (the source density
fluctuations) and $\tdelta_{\kappaclumpbar}$ (the self-shielded clump
opacity fluctuations) can be written as a bias (respectively
$\bu$, $b_j$ and $\bclump$) times the fiducial cosmic density field,
and further define an effective source bias
\begin{equation}
b_{j,\mathrm{eff}} = (1-\betaHI \,\betar)\,b_j - \betaclump \,\bclump + \betaHI \,\betar\, \bu\textrm{,}\label{eq:bjeff-def}
\end{equation}
which takes into account the recombination emission from the IGM and
absorption from the clumps\footnote{I will assume that $\bclump=\bu$,
  since both unshielded and clumped \hi are included in the estimate
  made in Appendix \ref{sec:estimate-b_0}; it could plausibly be the
  case that $\bclump$ in reality differs from $\bu$ if the distinction
  between phases is made carefully -- but since both $b_j$ and
  $\bclump$ only enter through $\bjeff$, any uncertainty in $\bclump$
  is degenerate with the uncertainty in $b_j$ which will be explored
  later.}. The intergalactic \hi density then follows immediately,
\begin{equation}
\hspace{0.6cm} 
\tdeltaHI = \frac{\left[\bu - \bjeff\, S(k)\right]\tdelta_{\rho} - \left[1-\betaHI\,
  \betar\right] S(k)\, \tdeltaSN }{1-\betaHI\, S(k)} \textrm{,}\label{eq:bHI-main}
\end{equation}
using equation \eqref{eq:j-decomposition}. The \hi density perturbation $\tdeltaHI$ can be split
into two terms, corresponding to the correlated and shot-noise
components respectively. The correlated part obeys $\tdeltaHI = \bHI
\tdelta_{\rho}$ where
\begin{equation}
\bHI(k) = \frac{\bu - \bjeff\, S(k)}{1-\betaHI\, S(k)}\textrm{,}\label{eq:bHI}
\end{equation}
showing that the \hi density traces the cosmological density in a
scale-dependent way.
This is the main result of the present
work. Its implications will be discussed in the next section.

\section{Results}\label{sec:results}

\subsection{Bias and power spectrum}

The preceding section used a systematic linearization of
first-principles radiative transfer to derive 
equations governing the IGM \hi density on large scales. I will now
explore what this implies for the bias $\bHI$ \eqref{eq:bHI} and total
power spectrum.  As previously discussed, a number of uncertain
parameters enter the calculation, namely: the IGM bias in the uniform
radiation limit, $\bu$; the effective source density $\bar{n}$; the
effective source bias $b_j$; the fraction of opacity in
collisionally-ionized clumps $\pclump$; and the mean physical
Lyman-limit opacity $\kappaHI = (\betaHI +\betaclump) \kappatb$. This
section will explore the consequences of varying all of these except for
$\bu = 1.5$ (see Appendix \ref{sec:estimate-b_0} for details). I will
explore substantial variations around the default choices (justified
earlier in the text) of $b_j=3$, \updated{$\bar{n}= 10^{-4}
h^3\,\Mpc^{-3}$}, $\pclump=0.10$ and $(a\kappaHI)^{-1} = 570\,\Mpc$
comoving.

\begin{figure}
\includegraphics[width=0.5\textwidth]{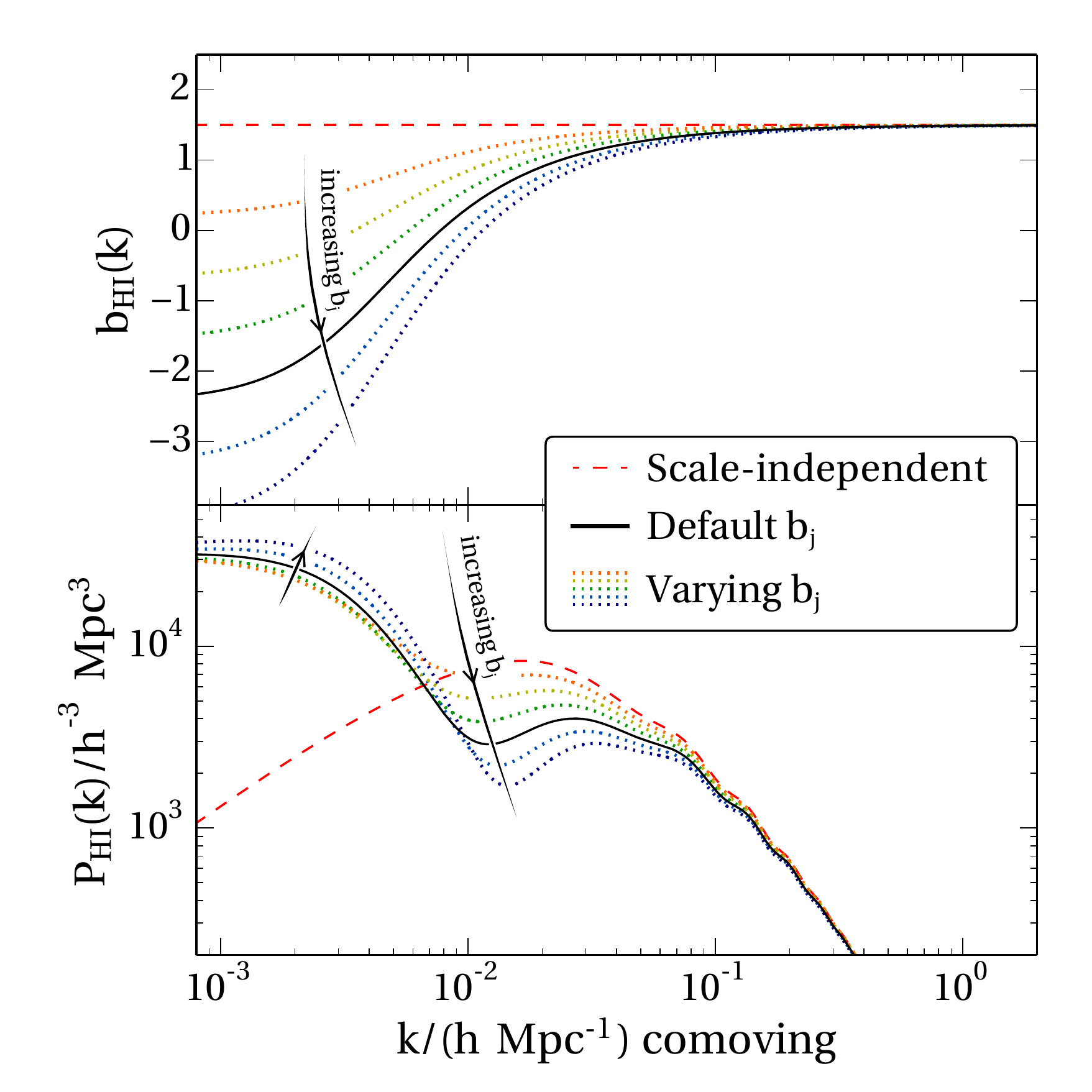}
\caption{({\it Upper panel}) the calculated bias of intergalactic \hi
  at $z=2.3$ as function of comoving wavenumber. The dashed line shows
  the bias in the uniform-radiation case; the solid line shows the bias calculated
  with radiative transfer. On small scales (towards the right), the
  calculated bias agrees with that of the uniform
  case. On large scales, the \hi is negatively biased because
  overdensities imply high emissivity, high radiation density and
  hence net \hi under-density.  Dotted lines show the effect of changing
  the source bias $b_j$; from top to bottom, $b_j=1.5$, $2.0$,
  $\cdots$, $4.0$. ({\it Lower panel}) the corresponding power
  spectrum, $P_{\mathrm{HI}}(k)$, defined by equation \eqref{eq:PHI},
  has a strong feature where $\bHI$ passes through zero (near to
  $k=\kappatb^{-1}$). }\label{fig:bk}
\end{figure}

The top panel of Figure \ref{fig:bk} plots $\bHI$ against comoving
wavenumber $k$, equation \eqref{eq:bHI}. This represents the linear relationship between \hi
density and total density as a function of scale. The dashed and solid
lines show respectively the assumed relationship when there are no
effects of inhomogeneous radiation and the calculated relationship
for the default parameters.

The basic functional form and its $b_j$ dependence can be understood
as follows. On small scales ($k \gg \kappatb$), $S(k)$ asymptotes to zero, so
\begin{equation}
\bHI(k \gg \kappatb) = \bu\textrm{,}
\end{equation}
showing that the small modes are unaffected by radiative transfer
phenomena at the linear level. Conversely on large scales, $S(k)$
asymptotes to one, giving
\begin{equation}
\bHI(k \ll \kappatb) = \frac{\bu-\bjeff }{1-\betaHI}\label{eq:large-limit}
\end{equation}
For $\bjeff>\bu$, this makes the \hi negatively biased on large
scales, i.e. anti-correlated with the total density. The intensity of
the radiation in dense regions over-compensates for the clustering
of hydrogen, causing a net deficit in neutral hydrogen -- a direct
analogue of the proximity effect but averaged over many sources on
large scales.

This has profound consequences for the power spectrum of \hi
fluctuations. Recall that $\tdeltaSN$ and $\tdelta_{\rho}$ are
uncorrelated, so we have
\begin{equation}
P_{\mathrm{HI}}(k) = b_{\mathrm{HI}}(k)^2 P(k) + \left[\frac{(1-\betaHI\,
  \betar)\,S(k)}{1-\betaHI S(k)}\right]^2 \frac{1}{\bar{n}}\textrm{.}\label{eq:PHI}
\end{equation}
This power spectrum is plotted in the lower panel of Figure
\ref{fig:bk}, with the default value $\bar{n} = 5 \times 10^{-4}
h^3\,\Mpc^{-3}$ (Section \ref{sec:emissivity}) and a fiducial $P(k)$
for the Planck cosmology calculated using CAMB
\cite{Lewis:1999bs}. The strong feature arises because $\bHI^2$
touches zero at $k^{-1} \approx 125 h^{-1} \,\Mpc $ comoving (for the
default parameters). Accordingly there is a sharp dip in $P(k)$ around
that wavenumber.  On larger scales still, at the far left of
Figure~\ref{fig:bk}, the \hi fluctuations become stronger than
predicted in the scale-independent model. This arises from a mixture
of shot-noise (discussed in more detail below) and the large
magnitude\footnote{Although $\bHI$ turns negative on large scales,
  this is not directly seen in the power spectrum which is sensitive
  only to $\bHI^2$. On the other hand $\bHI$ appears linearly when the
  forest is cross-correlated with another tracer population
  \cite{FontRibera_BOSS_QSO_Lya_X_2013,2012FontRibera_BOSS_DLA_Forest_X},
  so its sign is detectable in principle, a point explored a little
  more in a moment.} of the limiting bias \eqref{eq:large-limit}.

Dotted lines in Figure \ref{fig:bk} explore the impact of changing
$b_j$ over a wide range; from top to bottom, $b_j=1.5$, $2.0$,
$\cdots$, $4.0$.  Recall that, as discussed in Section
\ref{sec:emissivity}, the source bias $b_j$ is composed of a
photon-weighted average of different populations (excluding
recombination emission, which is accounted for elsewhere within the
calculation). As the source bias increases, the effects at a given
wavenumber typically become stronger. Consequently the zero in $\bHI$
moves to larger wavenumbers (shorter distances), making the radiation thumbprint
more observationally accessible. Even for small biases, however
($b_j=1.5$) the effects are significant on scales of tens to hundreds
of megaparsecs comoving. 

On sufficiently small scales, the \hi power spectrum is unaffected by
radiative transfer, regardless of the value of $b_j$. In particular,
1D measurements of the Lyman alpha forest are limited by the small
path length that can be observed with an individual quasar. Only
wavenumbers greater than $\sim 0.02 s\,\km^{-1}$, corresponding to
$2\,h\,\Mpc^{-1}$ comoving at $z=2.3$, are measured
\cite{McDonald_SDSS_1DLyA_Pk_2006}; from Figure \ref{fig:bk} it is
clear that the effects are minimal for such measurements. (The
non-linear, non-Gaussian contribution to the shot noise on those
scales will become significant, but plausibly average out over many
sightlines \cite{Slosar09BAO,White10_LyA_sims}.)

\subsection{Auto- and cross-correlation function}

To test the large-scale predictions against observations one must turn to more recent 3D
analyses that take advantage of modern surveys with dense background
sources \cite{2013Dawson_BOSS_overview}. In this context, it is more
conventional to consider the correlation function
$\xi(\vec{r}) = \langle
\delta(\vec{x}+\vec{r}) 
\delta(\vec{x})\rangle$. It has been widely used to constrain
the BAO feature in Lyman-$\alpha$ and other large scale structure
tracers
\cite{2013_Busca_BOSS_Lya_BAO,BOSS_Slosar13,BOSS_Anderson13_BAOGals}. By
isotropy $\xi$ is actually a
function of $r=|\vec{r}|$ alone; one can show it is related to the
power spectrum via a Legendre transformation,
\begin{equation}
\xi_{\mathrm{HI}}(r) = \frac{1}{2 \pi^2} \int \dd k\, \frac{\sin kr }{ kr} k^2 P_{\mathrm{HI}}(k)\textrm{.}\label{eq:legendre}
\end{equation} 
The correlation function for the Lyman-$\alpha$ flux $\xi_F$ is
closely related to $\xi_{\mathrm{HI}}$ and can be measured from
observations relatively directly. As
explained in the introduction, this paper will not go as far as
calculating $\xi_F$, but a brief discussion of the relationship to
$\xi_{\mathrm{HI}}$ is given in Section \ref{sec:discussion}.

\begin{figure}
\includegraphics[width=0.5\textwidth]{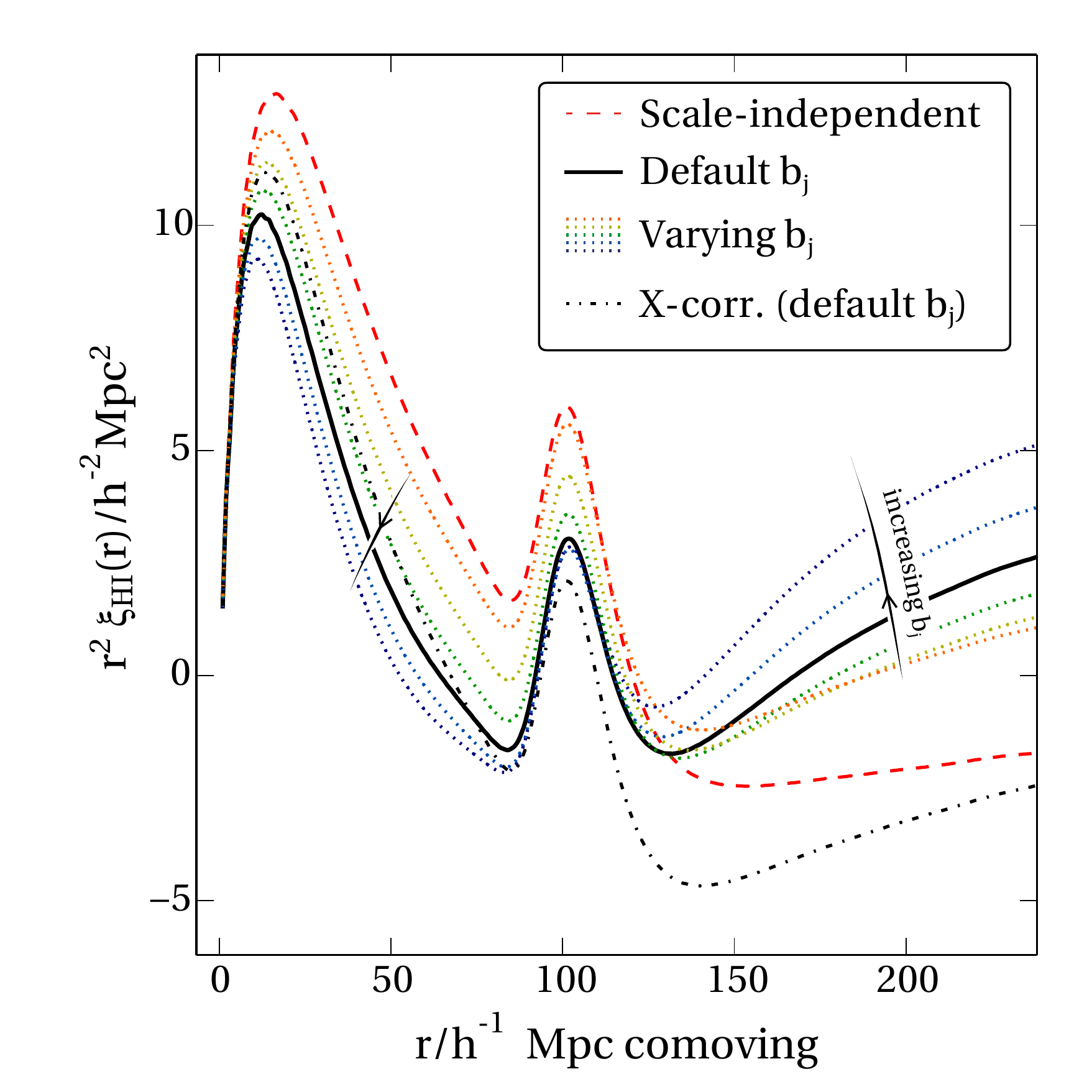}
\caption{The correlation function of intergalactic \hi at $z=2.3$, as
  defined by the Legendre transform~\protect\eqref{eq:legendre} of the
  power spectrum shown in Figure \ref{fig:bk}. As before, dashed lines
  show the constant bias case, whereas the solid line shows the
  calculated bias for inhomogeneous radiation (in the default case,
  $b_j=3.0$).  The dotted lines show a series of different source
  biases ($b_j=1.5$, nearest the dashed line; then $2.0$, $2.5$, $3.5$
  and $4.0$). The result of cross-correlating the \hi against a tracer
  population with fixed bias is shown by the dash-dotted
  line.}\label{fig:xi}
\end{figure}

Figure \ref{fig:xi} shows $\xi_{\mathrm{HI}}$ (renormalized by $r^2$
to highlight the BAO structure) for the scale-free (dashed line) and
default radiation model (solid line).  Once again the dotted lines
show the calculated correlation function for a range of different
source biases $b_j$ from $1.5$ to $4.0$. The mapping from power
spectrum to correlation function causes a substantial mixing of
information on different scales, so the new shape needs a little
unpicking to understand. On scales smaller than $\sim
5\,h^{-1}\,\Mpc$, the scale-free predictions are barely altered; this
corresponds to the small-scale limit $\bHI \to \bu$ in the bias,
Figure \ref{fig:bk}. Moving to larger separations, the
radiation-corrected correlation function falls rapidly compared to the
scale-free counterpart, because the \hi bias is declining and the
power on these scales is suppressed. In fact, the new correlation
function turns negative at around $55\,h^{-1}\,\Mpc$; this is an
artifact of the constraint that $\int \xi(r) r^2 \dd r=0$ for a
properly mean-calibrated sample, and the negativity in itself does not
indicate anything physically special about these scales.

\begin{figure}
\includegraphics[width=0.5\textwidth]{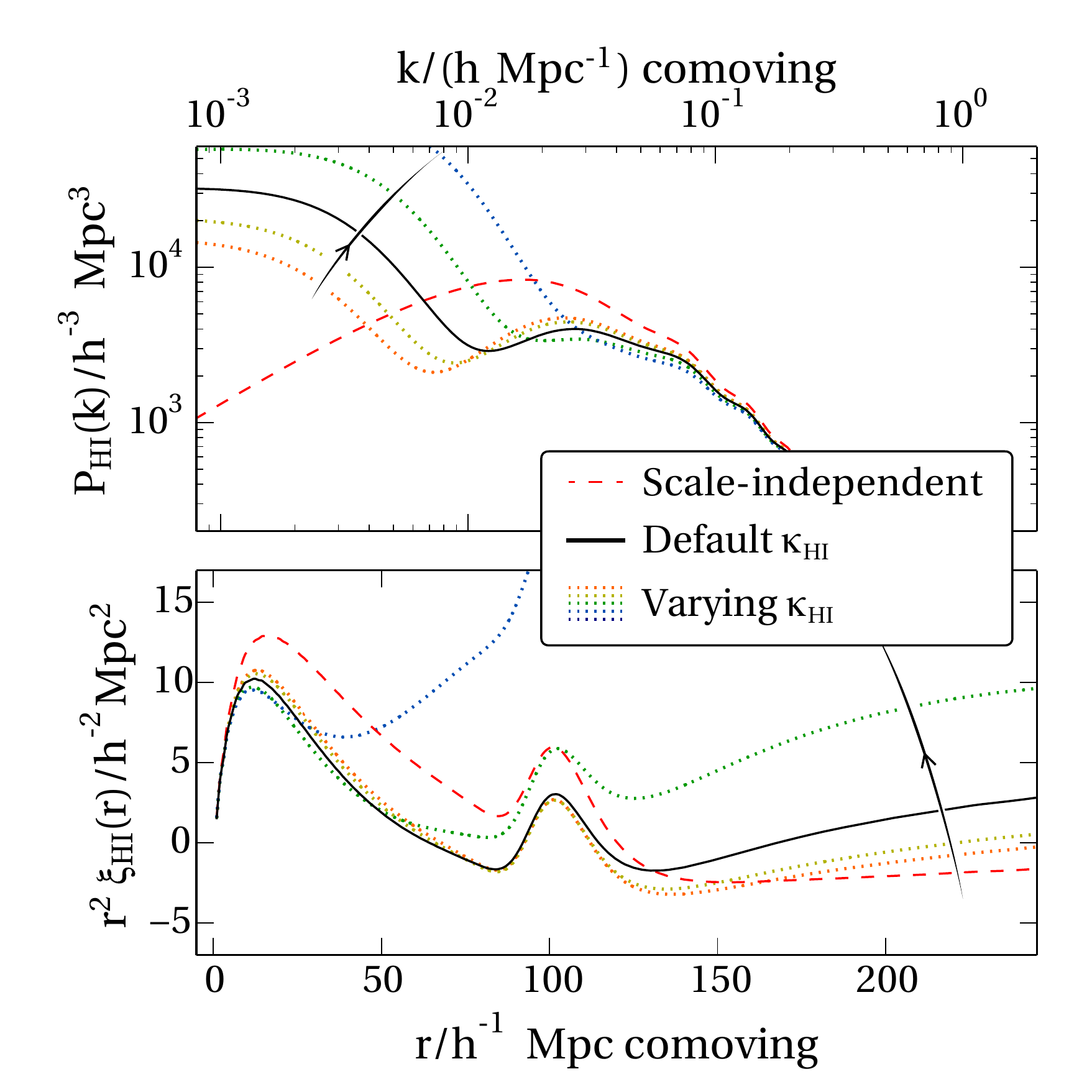}
\caption{The effect of varying the mean \hi opacity on the power
  spectrum (upper panel) and correlation function (lower panel). Other
  parameters are held fixed.  The dashed and solid lines show the
  uniform-radiation and reference case respectively, so agreeing with
  the same lines in Figures \ref{fig:bk} and \ref{fig:xi}. Dotted
  lines show the results for an \hi opacity $0.25$, $0.5$, $2$ and $4$
  times that of the default case. As the opacity increases, the
  mean-free-path decreases, meaning that the ``dip'' feature in the
  \hi power spectrum moves to shorter wavenumbers. Consequently
  small-scale power is increasingly suppressed, whereas large-scale
  power is enhanced.}\label{fig:mfp}
\end{figure}

\begin{figure}
\includegraphics[width=0.5\textwidth]{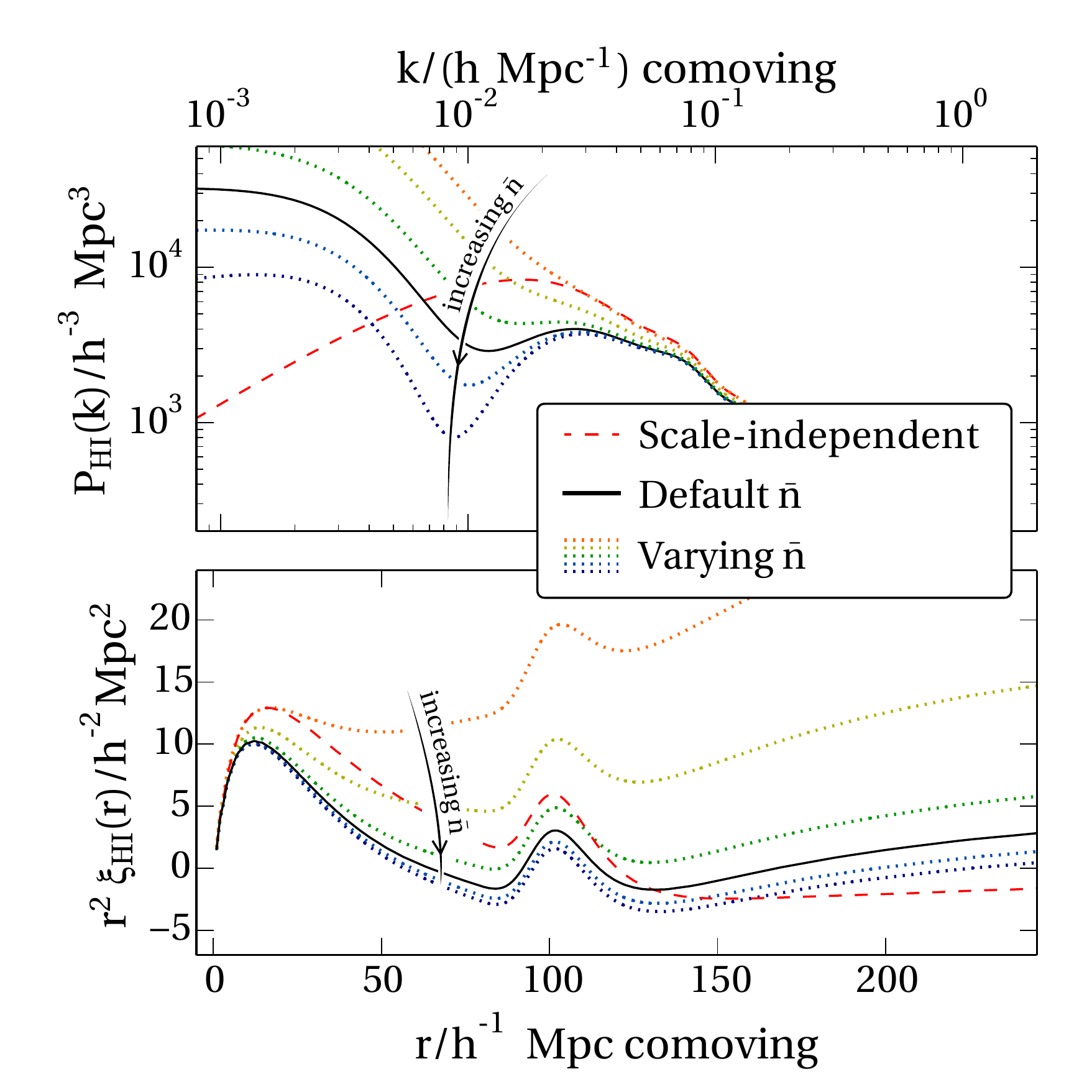}
\caption{The effect of source shot-noise on the power spectrum (upper
  panel) and correlation function (lower panel). Other parameters are
  held fixed. The dashed and solid lines show the uniform-radiation
  and reference case respectively, so agreeing with the same lines in
  Figures \ref{fig:bk} and \ref{fig:xi}. The dotted lines show the
  results for (top to bottom) $\bar{n}=10^{-5}$, $5 \times
  10^{-5}$, $10^{-4}$, $5 \times 10^{-4}$ (solid
  line), $10^{-3}$ and $5\times 10^{-3}\,h^3\,\Mpc^{-3}$. As the effective source
  density $\bar{n}$ decreases, the amplitude of shot-noise increases,
  confusing the \hi signal. The changes are strongest on large scales
  for the reasons discussed in the text. }\label{fig:shotnoise}
\end{figure}

The BAO feature -- a hump at around $r=100\,h^{-1}\,\Mpc$ -- remains
clearly visible in all cases, but the local maximum in $r^2 \xi(r)$
shifts marginally. The local maximum can be found at
$100.0\,h^{-1}\,\Mpc$ in the homogeneous case (dashed line) but at
$101.2\,h^{-1}\,\Mpc$ in the fiducial $b_j=3.0$ case (solid line). At
distances exceeding $130\,h^{-1}\,\Mpc$, the new correlation function
begins to rise because of contributions from the increased power on
very large scales (far left of Figure \ref{fig:bk}).

In cross-correlation, the signature looks slightly different. As an
illustration, the dot-dashed line in Figure \ref{fig:xi} shows a
hypothetical cross-correlation against a fixed-bias population with
$b=1.5$ (this value has no significance except to scale the overall
function similarly to the auto-correlation). In other words, I am
plotting $\xi_{\mathrm{HI} \times} \equiv 1.5\langle
\delta_{\mathrm{HI}}(\vec{x}+\vec{r})
\delta_{\rho}(\vec{x})\rangle$. In simple cases this would return the
geometric mean of the dashed and solid lines. However, there are a
couple of more subtle effects here. First, the negativity of the \hi
bias on large scales reduces the large-distance cross-correlation
($\xi_{\mathrm{HI}\times}$ probes $\bHI$ whereas $\xi_{\mathrm{HI}}$
is sensitive only to $\bHI^2$). Second, the plot assumes
cross-correlation against a population other than quasars so that the
shot-noise term cancels. Overall this leads to a cross-correlation
that is suppressed more strongly on large scales than would be
expected from an averaging argument.

\subsection{Varying other parameters}

So far I have only shown results for varying $b_j$. However there are
other parametric dependencies which ought to be examined. The first is
the physical \hi opacity, $\kappaHI$, which is the inverse of the
mean-free-path of a photon in the absence of redshifting or
volume-dilution. The default value has been discussed extensively
above; in Figure \ref{fig:mfp} I have shown what happens when
$\kappaHI$ is changed by a factor of $0.25$, $0.5$, $2$ and~$4$. The
actual uncertainty in the observationally-constrained value
\cite{Rudie13MFP,Prochaska13_MFP} is more likely under a factor of
2. The upper and lower panels show the power spectrum and correlation
function respectively. Solid and dashed lines therefore correspond
exactly to those presented in Figures \ref{fig:bk} and \ref{fig:xi};
the dotted lines show the impact of changing $\kappaHI$. As this
opacity increases (i.e. the mean-free-path decreases),
the effects becomes more prominent on smaller scales. A slightly
more subtle change occurs at long wavelengths: as the mean-free-path
decreases, the large-scale limiting bias increases, as does the noise
contribution. Since $\betaHI$ increases when $H$ is fixed but
$\kappaHI$ increases, this behaviour is in accordance with equation
\eqref{eq:large-limit}. Physically, photo-ionized \hi amplifies
fluctuations in radiation on large scales: an overdensity of radiation
implies a lower \hi fraction and therefore a deficit in opacity, in
turn boosting the overdensity of radiation. This is why, as $\kappaHI$
increases, the fluctuations on large scales become more dramatic.

Next consider the effect of varying $\bar{n}$ from its fiducial
value. Recall that this determines the large-scale shot-noise
contribution and is related to the underlying source population
densities (Section \ref{sec:emissivity}). Fixing the other parameters,
Figure \ref{fig:shotnoise} demonstrates the effect of $\bar{n}$
varying between $5\times 10^{-3}\, h^3\,\Mpc^{-3}$ and $10^{-5}\,
h^3\,\Mpc^{-3}$ on the power spectrum (upper panel) and
auto-correlation function (lower panel). \updated{Smaller source densities lead
to a stronger effect, with significant power added in the case of
$\bar{n}=10^{-5}\,h^3\,\Mpc^{-3}$. It may come as a
surprise that, in all cases,} the effects of low source density are most
pronounced as $r$ becomes large (or $k$ small) rather than in the
opposite limit. In the linear, averaged limit, however, this is
correct.  The shot-noise power spectrum is suppressed on small scales
by $S(k)^2$ which declines steeply at increasing $k$ (Figure
\ref{fig:Sk}). The intuitive picture that shot-noise matters more on
small scales depends on the transition to the non-linear, unaveraged
regime which I have not attempted to model.

\begin{figure}
\includegraphics[width=0.5\textwidth]{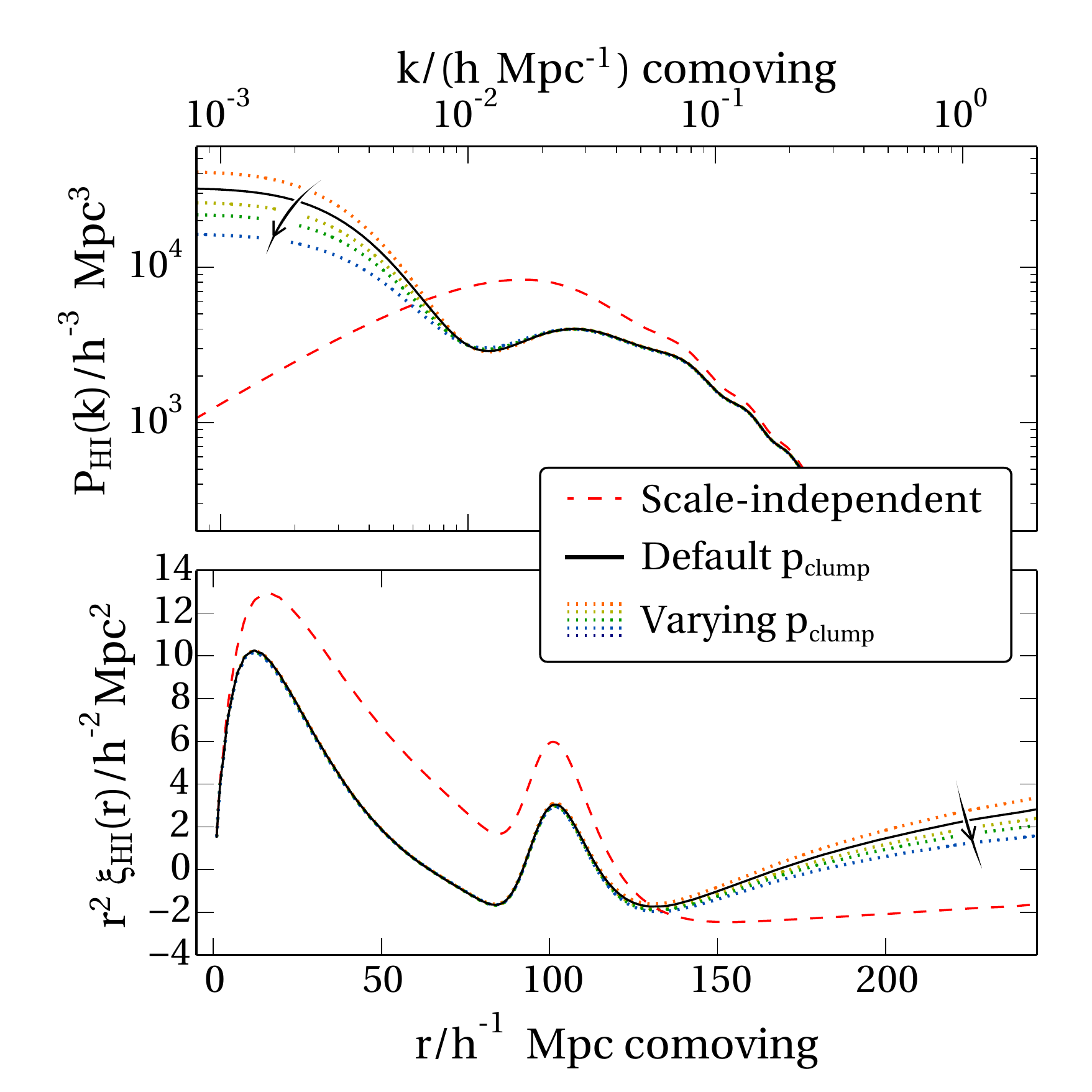}
\caption{The effect of changing the fraction of opacity from
  collisionally-ionized clumps on the power spectrum (upper panel) and
  correlation function (lower panel). Other parameters are held
  fixed. The dashed and solid lines show the uniform-radiation and
  reference case respectively, so agreeing with the same lines in
  Figures \ref{fig:bk} and \ref{fig:xi}. The dotted lines show the
  results for (top to bottom) $\pclump=0.0$, $0.1$ (solid line),
  $0.2$, $\cdots$, $0.5$. When more opacity arises from clumps (higher
  $\pclump$), the effect of radiation is slightly weaker because the
  clumps are able to partially counterbalance the enhanced radiation
  in overdense regions. However, the effect is minor.}\label{fig:clumping}
\end{figure}

Finally let us return to the parameter $\pclump$, which controls the
fraction of opacity arising from self-shielded, collisionally-ionized
clumps as opposed to diffuse, photo-ionized \hi. As this parameter is
increased, $\betaHI$ decreases and $\betaclump$ increases. The overall
effects are shown in Figure \ref{fig:clumping} for $\pclump =
0.0$, $0.1$ (the default), $0.2$, $\cdots$, $0.5$. For scenarios with a
greater fraction of opacity in clumps, the effect of radiation is
slightly mitigated on very large scales. However the differences are
minor. 

For realistic observations, the effects of $\pclump$ will be somewhat
larger: here I am plotting the effect only on the photo-ionized
intergalactic medium. Increasing the fraction of clumps contributing
to the Lyman-limit opacity will also increase the balance of such
systems in the Lyman-$\alpha$ forest flux spectrum. Since they are
collisionally ionized, they are not much affected by the inhomogeneous
radiation field and therefore they dilute the scale-dependent effects
roughly by a fraction $1-\pclump$. In other words, the leading-order
effect of a large $\pclump$ on observations will be different to, and
more important than, the physical effect on the intergalactic \hi
which I have discussed here.
 
Nonetheless, whatever the value of any of these parameters, there are
substantial changes to the intergalactic \hi correlation function at
all scales exceeding $5\,h^{-1}\,\Mpc$. It seems likely that these
should be detectable with BOSS observations of the Lyman-$\alpha$
forest -- even if observational complications lead to a substantial
dilution.  This prospect is considered further in the discussion
below.

\section{Discussion}\label{sec:discussion}

Radiative transfer imprints dramatic scale-dependent bias in the
intergalactic \hi and therefore the Lyman-$\alpha$ forest, even after
reionization is complete. This follows because regions separated by
distances larger than the UV photon mean-free-path reach essentially
independent photoionization equilibria. Source clustering is stronger
than IGM clustering, leading to negative \hi bias on large scales
(i.e. the \hi anti-correlates with large-scale overdensities).

This paper has presented a detailed calculation of these new effects
by adopting a monochromatic, equilibrium, large-scale description,
focusing on the large-scale, average correlations
\cite{Croft04_LyA_large_radiation,McDonald05,McQuinn11_LyA_fluctuations}
rather than small-scale non-linear fluctuations
\cite{Viel_WDM_Lya_2008,Slosar09BAO}. The systematic analytic
treatment starts from first-principles radiative transfer and
produces, with minimal computational effort, predictions for very
large scales.

The calculation reveals, as expected from the argument above, a strong
feature in the \hi power spectrum and a corresponding distortion of
its correlation function. According to the estimates here, this
distortion should have an effect at the BAO scale ($\approx
100\,h^{-1}\,\Mpc$). The BAO bump position is slightly shifted -- in
Figure \ref{fig:xi}, the local maximum is at 1.2\% larger scales in
the radiative-transfer case (solid line) compared to the constant-bias
case (dashed line). That said, future cosmology constraints are
unlikely to come from measuring the peak in such a simple way; so long
as algorithms marginalize over possible broadband distortions to the
correlation function, they will likely still recover an unbiased
estimate of the BAO scale.

The most interesting conclusion is therefore that BAO-focused
Lyman-$\alpha$ observational programmes will be able to recover
helpful astrophysical constraints: the radiative transfer distortions
are strongly dependent on the mean bias of sources (Figures
\ref{fig:bk} and \ref{fig:xi}), the \hi opacity (Figure \ref{fig:mfp})
and the effective number density of sources $\bar{n}$ (Figure
\ref{fig:shotnoise}). As $b_j$ increases, the correlated component of
the radiative fluctuations grows and the power on BAO scales decreases
while the power on very large scales increases; as $\bar{n}$
decreases, the random component of the radiative fluctuations grows
and the power on all scales increases. These trends seem to agree with
numerical results where a comparison can be made
\cite{Croft04_LyA_large_radiation,McDonald05,McQuinn11_LyA_fluctuations}.

The overall picture gives rise to a large number of questions. The
most obvious is whether existing BOSS observations of the
Lyman-$\alpha$ forest are compatible with the expected thumbprint.  A
variety of subtle observational issues must be taken into account
before this can be answered. First, converting an \hi correlation
function into a flux correlation function is a non-linear process that
needs at a minimum to be calibrated by suitable numerical simulations
\cite{McDonald03}. Redshift-space distortions will mix the dynamical
growth of structure with the tracer statistics into a final observed
correlation function
\cite{1987MNRAS.227....1K,1999coph.book.....P,McDonald03}. Dependent
on the exact survey design, angular binning and data cuts, these
effects could easily dilute the scale-dependence, making the observed
correlation function closer to the homogeneous-radiation
result. However the changes in the underlying intergalactic \hi bias
are sufficiently dramatic for it to seem implausible that the
radiation-transfer signature would be obscured completely in
forthcoming precision data. \updated{To be sure we will have to
  understand how the data processing and parameter degeneracies
  impact on our ability to measure the effects. Although the
  distortion is large, it is also a very smooth function of scale and
  therefore one needs to accurately calibrate the normalization of the
  correlation function over a wide range of scales to make a
  definitive detection; otherwise the effects are degenerate with a
  renormalization of the uniform-limit bias $\bu$.}

The BOSS team have emphasized that their 3D Lyman-$\alpha$ forest
pipeline is presently designed to pick out localized
correlation-function features -- i.e. the BAO bump -- rather than
reconstruct the entire function accurately
\cite{2013_Busca_BOSS_Lya_BAO,BOSS_Slosar13}. Nonetheless an attempt
to measure scale-dependence in cross-correlation against quasars has
been made; none was found \cite{FontRibera_BOSS_QSO_Lya_X_2013}.
Conversely some scale-dependence in the cross-correlation between
damped Lyman alpha systems and the forest can be seen in the plots of
Ref.  \cite{2012FontRibera_BOSS_DLA_Forest_X}.  It is unclear whether
and how these results can be reconciled with the present work;
observational difficulties such as continuum determination cause
severely correlated errors in correlation functions and dealing with
these leads to certain large-scale modes being projected out
\cite{2013_Busca_BOSS_Lya_BAO,BOSS_Slosar13}. Overall, the task of
determining whether the effects of radiative transfer are present in
existing data is considerable. However, I hope that the present
calculation has underlined the rewards of such an effort. The
scale-dependent radiative transfer contains a rich, valuable source of
information on the nature of UV sources.

21cm emission studies will not be affected by these considerations
because the \hi they probe is largely in collisional- rather than
photo-ionization equilibrium \cite{PontzenDLA}. The 21cm absorption
forest would be affected in just the same way as the Lyman-$\alpha$
forest; but this phenomena is of most promise at high redshift before
or during reionization \cite{Pritchard21cm_review} -- so the present
calculation does not apply. \updated{One way to tackle the larger
  fluctuations at high redshift is to use a halo-model-based calculation
  \cite{Furlanetto04_reion_model}; alternatively, a linear theory
  approach has been taken to the problem by
  Ref. \cite{Zhang07_reion,Aloisio13_reion_fNL,Mao13_reion_fNL}. In
  these cases, an explicit time integration needs to be performed to
  follow the growth of ionized bubbles whereas in the present case the
  integration is absent because I have assumed equilibrium, making the
  present paper's calculations considerably simpler but more restricted in
  scope.  }

Depending on one's assumptions (for instance on the relative
importance of quasars to the UV background, and on the quasar
luminosity function), the rarity of sources also have a substantial
impact on very large scales. Here I have modeled the resulting shot
noise by a Gaussian approximation similar to that of
Ref. \cite{McQuinn11_LyA_fluctuations}; \updated{in that work, noise was
considered to be so large that the correlated component of the
radiation fluctuations was thrown out of the calculation. In the
present work the effects of shot-noise seem milder, which
reflects that I work at lower redshift (where the comoving density of
quasars has increased) and make greater allowance for a UV
contribution from star-forming galaxies.} Crucially, the correlated
component has a qualitatively different signature to the noise
component of the radiation field: the former reduces the power of \hi
fluctuations on large scales, whereas the latter can only ever add
power (at any scale). One effect that is absent from the present work
concerns scales below $10\,\Mpc$ or so -- here the noise would be
significantly amplified \cite{Mesinger09_LyA_fluctuating_reionisation}
by 2-halo and other nonlinear effects. Another missing aspect from my
analysis is that of time-dependence which could, for example, add
further confusion from quasar duty cycles.

With all this in mind it would be of great interest to supplement the
linear theory calculations of this work by revisiting the BAO-scale
correlation function of the Lyman-$\alpha$ forest using non-linear
numerical simulations of gigaparsec chunks of the IGM with correlated
sources, incorporating radiative transfer -- along the lines of work
described by Refs
\cite{Croft04_LyA_large_radiation,McDonald05,McQuinn11_LyA_fluctuations}.
Hints of the anti-correlation discussed at length in the present paper
have been seen before in such efforts
\cite{Croft04_LyA_large_radiation,White10_LyA_sims}. It would be
helpful to include large scale temperature fluctuations arising from
helium reionization \cite{Furlanetto09_He_fluc}. Or, one might be able
to tackle temperature fluctuations analytically by relaxing the monochromatic
assumption; it is worth re-emphasizing that the current work includes
the zero-order effects of hard photons (the spectral shape enters
through equation \eqref{eq:alphaz}), but not first-order changes from
local fluctuations in spectral shape. \updated{At this level of
  approximation, the gas thermal equilibrium is nearly unaffected by
  radiation intensity fluctuations \cite{HuiGnedin97_IGM_EoS} because
  the radiative heating rate can be approximately re-written as a
  function of density and temperature (via the ionization equilibrium
  condition). To answer the important question of how thermal
  fluctuations change the large-scale signal one therefore needs
  either to incorporate multi-wavelength radiative transfer or go
  beyond an equilibrium approximation -- or, preferably, both
  \cite{HuiGnedin97_IGM_EoS}.}

Further work is required to reach a unified view of how radiation
changes the observational prospects for cosmology and astrophysics
with the Lyman-$\alpha$ forest. The present investigation forms a first guide
to the effects that will dominate on the largest scales.

\section*{Acknowledgments}
I am grateful to the anonymous referee for insightful comments and
suggestions; to Jamie Bolton, Pedro Ferreira, Andreu Font, Martin
Haehnelt, David Marsh, Jordi Miralda Escud\'e, Philip Hopkins, An\v ze
Slosar, Hiranya Peiris and Matteo Viel for discussions; and to the
Royal Society for financial support. Some numerical results in this
paper were derived with the help of the pynbody framework
\cite{2013ascl.soft05002P}.  \bibliographystyle{revtex}
\bibliography{../bibtex/updates.bib,../bibtex/refs.bib}

\begin{thebibliography}{10}
\providecommand*{\bibinfo}[2]{#2}
\providecommand*{\eprint}[1]{#1}
\providecommand*{\url}[1]{#1}
\bibitem{Rauch98ARAA}
\bibinfo{author}{M.~{Rauch}}, \bibinfo{journal}{\araa}
  \bibinfo{volume}{\textbf{36}}, \bibinfo{pages}{267} (\bibinfo{date}{1998}),
  \eprint{astro-ph/9806286}.
\bibitem{Becker01Reionization}
\bibinfo{author}{R.~H. {Becker}}, \bibinfo{author}{X.~{Fan}},
  \bibinfo{author}{R.~L. {White}}, \bibinfo{author}{M.~A. {Strauss}},
  \bibinfo{author}{V.~K. {Narayanan}}, \bibinfo{author}{R.~H. {Lupton}},
  \bibinfo{author}{J.~E. {Gunn}}, \bibinfo{author}{J.~{Annis}},
  \bibinfo{author}{N.~A. {Bahcall}}, \bibinfo{author}{J.~{Brinkmann}},
  \emph{et~al.}, \bibinfo{journal}{\aj} \bibinfo{volume}{\textbf{122}},
  \bibinfo{pages}{2850} (\bibinfo{date}{Dec. 2001}), \eprint{astro-ph/0108097}.
\bibitem{Fan02Reionization}
\bibinfo{author}{X.~{Fan}}, \bibinfo{author}{V.~K. {Narayanan}},
  \bibinfo{author}{M.~A. {Strauss}}, \bibinfo{author}{R.~L. {White}},
  \bibinfo{author}{R.~H. {Becker}}, \bibinfo{author}{L.~{Pentericci}}, and
  \bibinfo{author}{H.-W. {Rix}}, \bibinfo{journal}{\aj}
  \bibinfo{volume}{\textbf{123}}, \bibinfo{pages}{1247} (\bibinfo{date}{Mar.
  2002}), \eprint{astro-ph/0111184}.
\bibitem{2006ARA&A..44..415F}
\bibinfo{author}{X.~{Fan}}, \bibinfo{author}{C.~L. {Carilli}}, and
  \bibinfo{author}{B.~{Keating}}, \bibinfo{journal}{\araa}
  \bibinfo{volume}{\textbf{44}}, \bibinfo{pages}{415} (\bibinfo{date}{Sep.
  2006}), \eprint{astro-ph/0602375}.
\bibitem{Croft02}
\bibinfo{author}{R.~A.~C. {Croft}}, \bibinfo{author}{D.~H. {Weinberg}},
  \bibinfo{author}{M.~{Bolte}}, \bibinfo{author}{S.~{Burles}},
  \bibinfo{author}{L.~{Hernquist}}, \bibinfo{author}{N.~{Katz}},
  \bibinfo{author}{D.~{Kirkman}}, and \bibinfo{author}{D.~{Tytler}},
  \bibinfo{journal}{\apj} \bibinfo{volume}{\textbf{581}}, \bibinfo{pages}{20}
  (\bibinfo{date}{Dec. 2002}), \eprint{astro-ph/0012324}.
\bibitem{McDonald03}
\bibinfo{author}{P.~{McDonald}}, \bibinfo{journal}{\apj}
  \bibinfo{volume}{\textbf{585}}, \bibinfo{pages}{34} (\bibinfo{date}{Mar.
  2003}), \eprint{astro-ph/0108064}.
\bibitem{BOSS_Slosar11}
\bibinfo{author}{A.~{Slosar}}, \bibinfo{author}{A.~{Font-Ribera}},
  \bibinfo{author}{M.~M. {Pieri}}, \bibinfo{author}{J.~{Rich}},
  \bibinfo{author}{J.-M. {Le Goff}}, \bibinfo{author}{{\'E}.~{Aubourg}},
  \bibinfo{author}{J.~{Brinkmann}}, \bibinfo{author}{N.~{Busca}},
  \bibinfo{author}{B.~{Carithers}}, \bibinfo{author}{R.~{Charlassier}},
  \emph{et~al.}, \bibinfo{journal}{\jcap} \bibinfo{volume}{\textbf{9}},
  \bibinfo{pages}{1}, \bibinfo{eid}{001} (\bibinfo{date}{Sep. 2011}),
  \eprint{1104.5244}.
\bibitem{Viel_WDM_Lya_2005}
\bibinfo{author}{M.~{Viel}}, \bibinfo{author}{J.~{Lesgourgues}},
  \bibinfo{author}{M.~G. {Haehnelt}}, \bibinfo{author}{S.~{Matarrese}}, and
  \bibinfo{author}{A.~{Riotto}}, \bibinfo{journal}{\prd}
  \bibinfo{volume}{\textbf{71}}(6), \bibinfo{pages}{063534},
  \bibinfo{eid}{063534} (\bibinfo{date}{Mar. 2005}), \eprint{astro-ph/0501562}.
\bibitem{Viel_WDM_Lya_2008}
\bibinfo{author}{M.~{Viel}}, \bibinfo{author}{G.~D. {Becker}},
  \bibinfo{author}{J.~S. {Bolton}}, \bibinfo{author}{M.~G. {Haehnelt}},
  \bibinfo{author}{M.~{Rauch}}, and \bibinfo{author}{W.~L.~W. {Sargent}},
  \bibinfo{journal}{Physical Review Letters} \bibinfo{volume}{\textbf{100}}(4),
  \bibinfo{pages}{041304}, \bibinfo{eid}{041304} (\bibinfo{date}{Feb. 2008}),
  \eprint{0709.0131}.
\bibitem{Boyarsky_WDM_Lya_2009}
\bibinfo{author}{A.~{Boyarsky}}, \bibinfo{author}{J.~{Lesgourgues}},
  \bibinfo{author}{O.~{Ruchayskiy}}, and \bibinfo{author}{M.~{Viel}},
  \bibinfo{journal}{\jcap} \bibinfo{volume}{\textbf{5}}, \bibinfo{pages}{12},
  \bibinfo{eid}{012} (\bibinfo{date}{May 2009}), \eprint{0812.0010}.
\bibitem{Slosar09BAO}
\bibinfo{author}{A.~{Slosar}}, \bibinfo{author}{S.~{Ho}},
  \bibinfo{author}{M.~{White}}, and \bibinfo{author}{T.~{Louis}},
  \bibinfo{journal}{\jcap} \bibinfo{volume}{\textbf{10}}, \bibinfo{pages}{19},
  \bibinfo{eid}{019} (\bibinfo{date}{Oct. 2009}), \eprint{0906.2414}.
\bibitem{BOSS_Slosar13}
\bibinfo{author}{A.~{Slosar}}, \bibinfo{author}{V.~{Ir{\v s}i{\v c}}},
  \bibinfo{author}{D.~{Kirkby}}, \bibinfo{author}{S.~{Bailey}},
  \bibinfo{author}{N.~G. {Busca}}, \bibinfo{author}{T.~{Delubac}},
  \bibinfo{author}{J.~{Rich}}, \bibinfo{author}{{\'E}.~{Aubourg}},
  \bibinfo{author}{J.~E. {Bautista}}, \bibinfo{author}{V.~{Bhardwaj}},
  \emph{et~al.}, \bibinfo{journal}{\jcap} \bibinfo{volume}{\textbf{4}},
  \bibinfo{pages}{26}, \bibinfo{eid}{026} (\bibinfo{date}{Apr. 2013}),
  \eprint{1301.3459}.
\bibitem{FontRibera_BOSS_QSO_Lya_X_BAO_2013}
\bibinfo{author}{A.~{Font-Ribera}}, \bibinfo{author}{D.~{Kirkby}},
  \bibinfo{author}{N.~{Busca}}, \bibinfo{author}{J.~{Miralda-Escud{\'e}}},
  \bibinfo{author}{N.~P. {Ross}}, \bibinfo{author}{A.~{Slosar}},
  \bibinfo{author}{{\'E}.~{Aubourg}}, \bibinfo{author}{S.~{Bailey}},
  \bibinfo{author}{V.~{Bhardwaj}}, \bibinfo{author}{J.~{Bautista}},
  \emph{et~al.}, \bibinfo{journal}{JCAP, submitted}  (\bibinfo{date}{Nov.
  2013}), \eprint{1311.1767}.
\bibitem{Becker07}
\bibinfo{author}{G.~D. {Becker}}, \bibinfo{author}{M.~{Rauch}}, and
  \bibinfo{author}{W.~L.~W. {Sargent}}, \bibinfo{journal}{\apj}
  \bibinfo{volume}{\textbf{662}}, \bibinfo{pages}{72} (\bibinfo{date}{Jun.
  2007}), \eprint{astro-ph/0607633}.
\bibitem{Bolton08}
\bibinfo{author}{J.~S. {Bolton}}, \bibinfo{author}{M.~{Viel}},
  \bibinfo{author}{T.-S. {Kim}}, \bibinfo{author}{M.~G. {Haehnelt}}, and
  \bibinfo{author}{R.~F. {Carswell}}, \bibinfo{journal}{\mnras}
  \bibinfo{volume}{\textbf{386}}, \bibinfo{pages}{1131} (\bibinfo{date}{May
  2008}), \eprint{0711.2064}.
\bibitem{Becker11HeIIreionization}
\bibinfo{author}{G.~D. {Becker}}, \bibinfo{author}{J.~S. {Bolton}},
  \bibinfo{author}{M.~G. {Haehnelt}}, and \bibinfo{author}{W.~L.~W. {Sargent}},
  \bibinfo{journal}{\mnras} \bibinfo{volume}{\textbf{410}},
  \bibinfo{pages}{1096} (\bibinfo{date}{Jan. 2011}), \eprint{1008.2622}.
\bibitem{Croft98}
\bibinfo{author}{R.~A.~C. {Croft}}, \bibinfo{author}{D.~H. {Weinberg}},
  \bibinfo{author}{N.~{Katz}}, and \bibinfo{author}{L.~{Hernquist}},
  \bibinfo{journal}{\apj} \bibinfo{volume}{\textbf{495}}, \bibinfo{pages}{44}
  (\bibinfo{date}{Mar. 1998}), \eprint{astro-ph/9708018}.
\bibitem{McDonald00}
\bibinfo{author}{P.~{McDonald}}, \bibinfo{author}{J.~{Miralda-Escud{\'e}}},
  \bibinfo{author}{M.~{Rauch}}, \bibinfo{author}{W.~L.~W. {Sargent}},
  \bibinfo{author}{T.~A. {Barlow}}, \bibinfo{author}{R.~{Cen}}, and
  \bibinfo{author}{J.~P. {Ostriker}}, \bibinfo{journal}{\apj}
  \bibinfo{volume}{\textbf{543}}, \bibinfo{pages}{1} (\bibinfo{date}{Nov.
  2000}), \eprint{astro-ph/9911196}.
\bibitem{Bolton13IGMtemp}
\bibinfo{author}{J.~S. {Bolton}}, \bibinfo{author}{G.~D. {Becker}},
  \bibinfo{author}{M.~G. {Haehnelt}}, and \bibinfo{author}{M.~{Viel}},
  \bibinfo{journal}{\mnras} \bibinfo{volume}{\textbf{438}},
  \bibinfo{pages}{2499} (\bibinfo{date}{Mar. 2014}), \eprint{1308.4411}.
\bibitem{Faucher-Giguere09_UVB}
\bibinfo{author}{C.-A. {Faucher-Gigu{\`e}re}}, \bibinfo{author}{A.~{Lidz}},
  \bibinfo{author}{M.~{Zaldarriaga}}, and \bibinfo{author}{L.~{Hernquist}},
  \bibinfo{journal}{\apj} \bibinfo{volume}{\textbf{703}}, \bibinfo{pages}{1416}
  (\bibinfo{date}{Oct. 2009}), \eprint{0901.4554}.
\bibitem{HM12}
\bibinfo{author}{F.~{Haardt}} and \bibinfo{author}{P.~{Madau}},
  \bibinfo{journal}{\apj} \bibinfo{volume}{\textbf{746}}, \bibinfo{pages}{125},
  \bibinfo{eid}{125} (\bibinfo{date}{Feb. 2012}), \eprint{1105.2039}.
\bibitem{Maselli05_LyA_transmissivity}
\bibinfo{author}{A.~{Maselli}} and \bibinfo{author}{A.~{Ferrara}},
  \bibinfo{journal}{\mnras} \bibinfo{volume}{\textbf{364}},
  \bibinfo{pages}{1429} (\bibinfo{date}{Dec. 2005}), \eprint{astro-ph/0510258}.
\bibitem{Zuo92}
\bibinfo{author}{L.~{Zuo}}, \bibinfo{journal}{\mnras}
  \bibinfo{volume}{\textbf{258}}, \bibinfo{pages}{36} (\bibinfo{date}{Sep.
  1992}).
\bibitem{Kollmeier03_LyA_LBG_connection}
\bibinfo{author}{J.~A. {Kollmeier}}, \bibinfo{author}{D.~H. {Weinberg}},
  \bibinfo{author}{R.~{Dav{\'e}}}, and \bibinfo{author}{N.~{Katz}},
  \bibinfo{journal}{\apj} \bibinfo{volume}{\textbf{594}}, \bibinfo{pages}{75}
  (\bibinfo{date}{Sep. 2003}), \eprint{astro-ph/0209563}.
\bibitem{Meiskin04_LyA_radiation}
\bibinfo{author}{A.~{Meiksin}} and \bibinfo{author}{M.~{White}},
  \bibinfo{journal}{\mnras} \bibinfo{volume}{\textbf{350}},
  \bibinfo{pages}{1107} (\bibinfo{date}{May 2004}), \eprint{astro-ph/0307289}.
\bibitem{Kollmeier06_LyA_galwinds}
\bibinfo{author}{J.~A. {Kollmeier}}, \bibinfo{author}{J.~{Miralda-Escud{\'e}}},
  \bibinfo{author}{R.~{Cen}}, and \bibinfo{author}{J.~P. {Ostriker}},
  \bibinfo{journal}{\apj} \bibinfo{volume}{\textbf{638}}, \bibinfo{pages}{52}
  (\bibinfo{date}{Feb. 2006}), \eprint{astro-ph/0503674}.
\bibitem{Mesinger09_LyA_fluctuating_reionisation}
\bibinfo{author}{A.~{Mesinger}} and \bibinfo{author}{S.~{Furlanetto}},
  \bibinfo{journal}{\mnras} \bibinfo{volume}{\textbf{400}},
  \bibinfo{pages}{1461} (\bibinfo{date}{Dec. 2009}), \eprint{0906.3020}.
\bibitem{White10_LyA_sims}
\bibinfo{author}{M.~{White}}, \bibinfo{author}{A.~{Pope}},
  \bibinfo{author}{J.~{Carlson}}, \bibinfo{author}{K.~{Heitmann}},
  \bibinfo{author}{S.~{Habib}}, \bibinfo{author}{P.~{Fasel}},
  \bibinfo{author}{D.~{Daniel}}, and \bibinfo{author}{Z.~{Lukic}},
  \bibinfo{journal}{\apj} \bibinfo{volume}{\textbf{713}}, \bibinfo{pages}{383}
  (\bibinfo{date}{Apr. 2010}), \eprint{0911.5341}.
\bibitem{FontRibera_BOSS_QSO_Lya_X_2013}
\bibinfo{author}{A.~{Font-Ribera}}, \bibinfo{author}{E.~{Arnau}},
  \bibinfo{author}{J.~{Miralda-Escud{\'e}}}, \bibinfo{author}{E.~{Rollinde}},
  \bibinfo{author}{J.~{Brinkmann}}, \bibinfo{author}{J.~R. {Brownstein}},
  \bibinfo{author}{K.-G. {Lee}}, \bibinfo{author}{A.~D. {Myers}},
  \bibinfo{author}{N.~{Palanque-Delabrouille}},
  \bibinfo{author}{I.~{P{\^a}ris}}, \emph{et~al.}, \bibinfo{journal}{\jcap}
  \bibinfo{volume}{\textbf{5}}, \bibinfo{pages}{18}, \bibinfo{eid}{018}
  (\bibinfo{date}{May 2013}), \eprint{1303.1937}.
\bibitem{Croft04_LyA_large_radiation}
\bibinfo{author}{R.~A.~C. {Croft}}, \bibinfo{journal}{\apj}
  \bibinfo{volume}{\textbf{610}}, \bibinfo{pages}{642} (\bibinfo{date}{Aug.
  2004}), \eprint{astro-ph/0310890}.
\bibitem{McDonald05}
\bibinfo{author}{P.~{McDonald}}, \bibinfo{author}{U.~{Seljak}},
  \bibinfo{author}{R.~{Cen}}, \bibinfo{author}{P.~{Bode}}, and
  \bibinfo{author}{J.~P. {Ostriker}}, \bibinfo{journal}{\mnras}
  \bibinfo{volume}{\textbf{360}}, \bibinfo{pages}{1471} (\bibinfo{date}{Jul.
  2005}), \eprint{astro-ph/0407378}.
\bibitem{McQuinn11_LyA_fluctuations}
\bibinfo{author}{M.~{McQuinn}}, \bibinfo{author}{L.~{Hernquist}},
  \bibinfo{author}{A.~{Lidz}}, and \bibinfo{author}{M.~{Zaldarriaga}},
  \bibinfo{journal}{\mnras} \bibinfo{volume}{\textbf{415}},
  \bibinfo{pages}{977} (\bibinfo{date}{Jul. 2011}), \eprint{1010.5250}.
\bibitem{Rudie13MFP}
\bibinfo{author}{G.~C. {Rudie}}, \bibinfo{author}{C.~C. {Steidel}},
  \bibinfo{author}{A.~E. {Shapley}}, and \bibinfo{author}{M.~{Pettini}},
  \bibinfo{journal}{\apj} \bibinfo{volume}{\textbf{769}}, \bibinfo{pages}{146},
  \bibinfo{eid}{146} (\bibinfo{date}{Jun. 2013}), \eprint{1304.6719}.
\bibitem{2013Dawson_BOSS_overview}
\bibinfo{author}{K.~S. {Dawson}}, \bibinfo{author}{D.~J. {Schlegel}},
  \bibinfo{author}{C.~P. {Ahn}}, \bibinfo{author}{S.~F. {Anderson}},
  \bibinfo{author}{{\'E}.~{Aubourg}}, \bibinfo{author}{S.~{Bailey}},
  \bibinfo{author}{R.~H. {Barkhouser}}, \bibinfo{author}{J.~E. {Bautista}},
  \bibinfo{author}{A.~{Beifiori}}, \bibinfo{author}{A.~A. {Berlind}},
  \emph{et~al.}, \bibinfo{journal}{\aj} \bibinfo{volume}{\textbf{145}},
  \bibinfo{pages}{10}, \bibinfo{eid}{10} (\bibinfo{date}{Jan. 2013}),
  \eprint{1208.0022}.
\bibitem{2013_Busca_BOSS_Lya_BAO}
\bibinfo{author}{N.~G. {Busca}}, \bibinfo{author}{T.~{Delubac}},
  \bibinfo{author}{J.~{Rich}}, \bibinfo{author}{S.~{Bailey}},
  \bibinfo{author}{A.~{Font-Ribera}}, \bibinfo{author}{D.~{Kirkby}},
  \bibinfo{author}{J.-M. {Le Goff}}, \bibinfo{author}{M.~M. {Pieri}},
  \bibinfo{author}{A.~{Slosar}}, \bibinfo{author}{{\'E}.~{Aubourg}},
  \emph{et~al.}, \bibinfo{journal}{\aap} \bibinfo{volume}{\textbf{552}},
  \bibinfo{pages}{A96}, \bibinfo{eid}{A96} (\bibinfo{date}{Apr. 2013}),
  \eprint{1211.2616}.
\bibitem{McQuinn_11_LyLimit}
\bibinfo{author}{M.~{McQuinn}}, \bibinfo{author}{S.~P. {Oh}}, and
  \bibinfo{author}{C.-A. {Faucher-Gigu{\`e}re}}, \bibinfo{journal}{\apj}
  \bibinfo{volume}{\textbf{743}}, \bibinfo{pages}{82}, \bibinfo{eid}{82}
  (\bibinfo{date}{Dec. 2011}), \eprint{1101.1964}.
\bibitem{Zhang07_reion}
\bibinfo{author}{J.~{Zhang}}, \bibinfo{author}{L.~{Hui}}, and
  \bibinfo{author}{Z.~{Haiman}}, \bibinfo{journal}{\mnras}
  \bibinfo{volume}{\textbf{375}}, \bibinfo{pages}{324} (\bibinfo{date}{Feb.
  2007}), \eprint{astro-ph/0607628}.
\bibitem{Planck13_CosmoPar}
\bibinfo{author}{{Planck Collaboration}}, \bibinfo{author}{P.~A.~R. {Ade}},
  \bibinfo{author}{N.~{Aghanim}}, \bibinfo{author}{C.~{Armitage-Caplan}},
  \bibinfo{author}{M.~{Arnaud}}, \bibinfo{author}{M.~{Ashdown}},
  \bibinfo{author}{F.~{Atrio-Barandela}}, \bibinfo{author}{J.~{Aumont}},
  \bibinfo{author}{C.~{Baccigalupi}}, \bibinfo{author}{A.~J. {Banday}},
  \emph{et~al.}, \bibinfo{journal}{\aap, submitted}  (\bibinfo{date}{Mar.
  2013}), \eprint{1303.5076}.
\bibitem{1996ApJ...461...20H}
\bibinfo{author}{F.~{Haardt}} and \bibinfo{author}{P.~{Madau}},
  \bibinfo{journal}{\apj} \bibinfo{volume}{\textbf{461}}, \bibinfo{pages}{20}
  (\bibinfo{date}{Apr. 1996}).
\bibitem{Prochaska13_MFP}
\bibinfo{author}{J.~X. {Prochaska}}, \bibinfo{author}{P.~{Madau}},
  \bibinfo{author}{J.~M. {O'Meara}}, and \bibinfo{author}{M.~{Fumagalli}},
  \bibinfo{journal}{\mnras} \bibinfo{volume}{\textbf{438}},
  \bibinfo{pages}{476} (\bibinfo{date}{Feb. 2014}), \eprint{1310.0052}.
\bibitem{PontzenDLA}
\bibinfo{author}{A.~{Pontzen}}, \bibinfo{author}{F.~{Governato}},
  \bibinfo{author}{M.~{Pettini}}, \bibinfo{author}{C.~M. {Booth}},
  \bibinfo{author}{G.~{Stinson}}, \bibinfo{author}{J.~{Wadsley}},
  \bibinfo{author}{A.~{Brooks}}, \bibinfo{author}{T.~{Quinn}}, and
  \bibinfo{author}{M.~{Haehnelt}}, \bibinfo{journal}{\mnras, accepted}
  (\bibinfo{date}{2008}).
\bibitem{Black81clouds}
\bibinfo{author}{J.~H. {Black}}, \bibinfo{journal}{\mnras}
  \bibinfo{volume}{\textbf{197}}, \bibinfo{pages}{553} (\bibinfo{date}{Nov.
  1981}).
\bibitem{loeb2013firstgals}
\bibinfo{author}{A.~Loeb} and \bibinfo{author}{S.~Furlanetto},
  \bibinfo{title}{\emph{The First Galaxies in the Universe}},
  \bibinfo{volume}{Princeton Series in Astrophysics}
  (\bibinfo{publisher}{Princeton University Press}, \bibinfo{year}{2013}), ISBN
  \bibinfo{isbn}{9781400845606}.
\bibitem{Lai05_LyA_temp_flucs}
\bibinfo{author}{K.~{Lai}}, \bibinfo{author}{A.~{Lidz}},
  \bibinfo{author}{L.~{Hernquist}}, and \bibinfo{author}{M.~{Zaldarriaga}},
  \bibinfo{journal}{\apj} \bibinfo{volume}{\textbf{644}}, \bibinfo{pages}{61}
  (\bibinfo{date}{Jun. 2006}), \eprint{astro-ph/0510841}.
\bibitem{Bolton06_UV_shape_fluctuations}
\bibinfo{author}{J.~S. {Bolton}}, \bibinfo{author}{M.~G. {Haehnelt}},
  \bibinfo{author}{M.~{Viel}}, and \bibinfo{author}{R.~F. {Carswell}},
  \bibinfo{journal}{\mnras} \bibinfo{volume}{\textbf{366}},
  \bibinfo{pages}{1378} (\bibinfo{date}{Mar. 2006}), \eprint{astro-ph/0508201}.
\bibitem{Furlanetto09_He_fluc}
\bibinfo{author}{S.~R. {Furlanetto}}, \bibinfo{journal}{\apj}
  \bibinfo{volume}{\textbf{703}}, \bibinfo{pages}{702} (\bibinfo{date}{Sep.
  2009}), \eprint{0812.3411}.
\bibitem{HuiGnedin97_IGM_EoS}
\bibinfo{author}{L.~{Hui}} and \bibinfo{author}{N.~Y. {Gnedin}},
  \bibinfo{journal}{\mnras} \bibinfo{volume}{\textbf{292}}, \bibinfo{pages}{27}
  (\bibinfo{date}{Nov. 1997}), \eprint{astro-ph/9612232}.
\bibitem{2011PhRvD..84f3505B}
\bibinfo{author}{C.~{Bonvin}} and \bibinfo{author}{R.~{Durrer}},
  \bibinfo{journal}{\prd} \bibinfo{volume}{\textbf{84}}(6),
  \bibinfo{pages}{063505}, \bibinfo{eid}{063505} (\bibinfo{date}{Sep. 2011}),
  \eprint{1105.5280}.
\bibitem{Croom2dF_QSObias05}
\bibinfo{author}{S.~M. {Croom}}, \bibinfo{author}{B.~J. {Boyle}},
  \bibinfo{author}{T.~{Shanks}}, \bibinfo{author}{R.~J. {Smith}},
  \bibinfo{author}{L.~{Miller}}, \bibinfo{author}{P.~J. {Outram}},
  \bibinfo{author}{N.~S. {Loaring}}, \bibinfo{author}{F.~{Hoyle}}, and
  \bibinfo{author}{J.~{da {\^A}ngela}}, \bibinfo{journal}{\mnras}
  \bibinfo{volume}{\textbf{356}}, \bibinfo{pages}{415} (\bibinfo{date}{Jan.
  2005}), \eprint{astro-ph/0409314}.
\bibitem{White_QSO_clustering}
\bibinfo{author}{M.~{White}}, \bibinfo{author}{A.~D. {Myers}},
  \bibinfo{author}{N.~P. {Ross}}, \bibinfo{author}{D.~J. {Schlegel}},
  \bibinfo{author}{J.~F. {Hennawi}}, \bibinfo{author}{Y.~{Shen}},
  \bibinfo{author}{I.~{McGreer}}, \bibinfo{author}{M.~A. {Strauss}},
  \bibinfo{author}{A.~S. {Bolton}}, \bibinfo{author}{J.~{Bovy}}, \emph{et~al.},
  \bibinfo{journal}{\mnras} \bibinfo{volume}{\textbf{424}},
  \bibinfo{pages}{933} (\bibinfo{date}{Aug. 2012}), \eprint{1203.5306}.
\bibitem{Adelberger05_galaxy_correlation}
\bibinfo{author}{K.~L. {Adelberger}}, \bibinfo{author}{C.~C. {Steidel}},
  \bibinfo{author}{M.~{Pettini}}, \bibinfo{author}{A.~E. {Shapley}},
  \bibinfo{author}{N.~A. {Reddy}}, and \bibinfo{author}{D.~K. {Erb}},
  \bibinfo{journal}{\apj} \bibinfo{volume}{\textbf{619}}, \bibinfo{pages}{697}
  (\bibinfo{date}{Feb. 2005}), \eprint{astro-ph/0410165}.
\bibitem{Reddy09_UV_LF}
\bibinfo{author}{N.~A. {Reddy}} and \bibinfo{author}{C.~C. {Steidel}},
  \bibinfo{journal}{\apj} \bibinfo{volume}{\textbf{692}}, \bibinfo{pages}{778}
  (\bibinfo{date}{Feb. 2009}), \eprint{0810.2788}.
\bibitem{ColeKaiser_Bias_89}
\bibinfo{author}{S.~{Cole}} and \bibinfo{author}{N.~{Kaiser}},
  \bibinfo{journal}{\mnras} \bibinfo{volume}{\textbf{237}},
  \bibinfo{pages}{1127} (\bibinfo{date}{Apr. 1989}).
\bibitem{2013ascl.soft05002P}
\bibinfo{author}{A.~{Pontzen}}, \bibinfo{author}{R.~{Roskar}},
  \bibinfo{author}{G.~{Stinson}}, and \bibinfo{author}{R.~{Woods}},
  \bibinfo{title}{\emph{{pynbody: N-Body/SPH analysis for python}}}
  (\bibinfo{date}{May 2013}), astrophysics Source Code Library,
  \eprint{ascl:1305.002}.
\bibitem{Moster13}
\bibinfo{author}{B.~P. {Moster}}, \bibinfo{author}{T.~{Naab}}, and
  \bibinfo{author}{S.~D.~M. {White}}, \bibinfo{journal}{\mnras}
  \bibinfo{volume}{\textbf{428}}, \bibinfo{pages}{3121} (\bibinfo{date}{Feb.
  2013}), \eprint{1205.5807}.
\bibitem{2007ApJ...654..731H}
\bibinfo{author}{P.~F. {Hopkins}}, \bibinfo{author}{G.~T. {Richards}}, and
  \bibinfo{author}{L.~{Hernquist}}, \bibinfo{journal}{\apj}
  \bibinfo{volume}{\textbf{654}}, \bibinfo{pages}{731} (\bibinfo{date}{Jan.
  2007}), \eprint{astro-ph/0605678}.
\bibitem{Ross13_SDSS9_QLF}
\bibinfo{author}{N.~P. {Ross}}, \bibinfo{author}{I.~D. {McGreer}},
  \bibinfo{author}{M.~{White}}, \bibinfo{author}{G.~T. {Richards}},
  \bibinfo{author}{A.~D. {Myers}},
  \bibinfo{author}{N.~{Palanque-Delabrouille}}, \bibinfo{author}{M.~A.
  {Strauss}}, \bibinfo{author}{S.~F. {Anderson}}, \bibinfo{author}{Y.~{Shen}},
  \bibinfo{author}{W.~N. {Brandt}}, \emph{et~al.}, \bibinfo{journal}{\apj}
  \bibinfo{volume}{\textbf{773}}, \bibinfo{pages}{14}, \bibinfo{eid}{14}
  (\bibinfo{date}{Aug. 2013}), \eprint{1210.6389}.
\bibitem{Lewis:1999bs}
\bibinfo{author}{A.~Lewis}, \bibinfo{author}{A.~Challinor}, and
  \bibinfo{author}{A.~Lasenby}, \bibinfo{journal}{Astrophys. J.}
  \bibinfo{volume}{\textbf{538}}, \bibinfo{pages}{473} (\bibinfo{date}{2000}),
  \eprint{astro-ph/9911177}.
\bibitem{2012FontRibera_BOSS_DLA_Forest_X}
\bibinfo{author}{A.~{Font-Ribera}}, \bibinfo{author}{J.~{Miralda-Escud{\'e}}},
  \bibinfo{author}{E.~{Arnau}}, \bibinfo{author}{B.~{Carithers}},
  \bibinfo{author}{K.-G. {Lee}}, \bibinfo{author}{P.~{Noterdaeme}},
  \bibinfo{author}{I.~{P{\^a}ris}}, \bibinfo{author}{P.~{Petitjean}},
  \bibinfo{author}{J.~{Rich}}, \bibinfo{author}{E.~{Rollinde}}, \emph{et~al.},
  \bibinfo{journal}{\jcap} \bibinfo{volume}{\textbf{11}}, \bibinfo{pages}{59},
  \bibinfo{eid}{059} (\bibinfo{date}{Nov. 2012}), \eprint{1209.4596}.
\bibitem{McDonald_SDSS_1DLyA_Pk_2006}
\bibinfo{author}{P.~{McDonald}}, \bibinfo{author}{U.~{Seljak}},
  \bibinfo{author}{S.~{Burles}}, \bibinfo{author}{D.~J. {Schlegel}},
  \bibinfo{author}{D.~H. {Weinberg}}, \bibinfo{author}{R.~{Cen}},
  \bibinfo{author}{D.~{Shih}}, \bibinfo{author}{J.~{Schaye}},
  \bibinfo{author}{D.~P. {Schneider}}, \bibinfo{author}{N.~A. {Bahcall}},
  \emph{et~al.}, \bibinfo{journal}{\apjs} \bibinfo{volume}{\textbf{163}},
  \bibinfo{pages}{80} (\bibinfo{date}{Mar. 2006}), \eprint{astro-ph/0405013}.
\bibitem{BOSS_Anderson13_BAOGals}
\bibinfo{author}{L.~{Anderson}}, \bibinfo{author}{E.~{Aubourg}},
  \bibinfo{author}{S.~{Bailey}}, \bibinfo{author}{F.~{Beutler}},
  \bibinfo{author}{A.~S. {Bolton}}, \bibinfo{author}{J.~{Brinkmann}},
  \bibinfo{author}{J.~R. {Brownstein}}, \bibinfo{author}{C.-H. {Chuang}},
  \bibinfo{author}{A.~J. {Cuesta}}, \bibinfo{author}{K.~S. {Dawson}},
  \emph{et~al.}, \bibinfo{journal}{\mnras} \bibinfo{volume}{\textbf{439}},
  \bibinfo{pages}{83} (\bibinfo{date}{Mar. 2014}), \eprint{1303.4666}.
\bibitem{1987MNRAS.227....1K}
\bibinfo{author}{N.~{Kaiser}}, \bibinfo{journal}{\mnras}
  \bibinfo{volume}{\textbf{227}}, \bibinfo{pages}{1} (\bibinfo{date}{Jul.
  1987}).
\bibitem{1999coph.book.....P}
\bibinfo{author}{J.~A. {Peacock}}, \bibinfo{title}{\emph{{Cosmological
  physics}}} (\bibinfo{publisher}{Cambridge University Press}, {Cambridge, UK},
  \bibinfo{year}{1999}).
\bibitem{Pritchard21cm_review}
\bibinfo{author}{J.~R. {Pritchard}} and \bibinfo{author}{A.~{Loeb}},
  \bibinfo{journal}{Reports on Progress in Physics}
  \bibinfo{volume}{\textbf{75}}(8), \bibinfo{pages}{086901},
  \bibinfo{eid}{086901} (\bibinfo{date}{Aug. 2012}), \eprint{1109.6012}.
\bibitem{Furlanetto04_reion_model}
\bibinfo{author}{S.~R. {Furlanetto}}, \bibinfo{author}{M.~{Zaldarriaga}}, and
  \bibinfo{author}{L.~{Hernquist}}, \bibinfo{journal}{\apj}
  \bibinfo{volume}{\textbf{613}}, \bibinfo{pages}{1} (\bibinfo{date}{Sep.
  2004}), \eprint{astro-ph/0403697}.
\bibitem{Aloisio13_reion_fNL}
\bibinfo{author}{A.~{D'Aloisio}}, \bibinfo{author}{J.~{Zhang}},
  \bibinfo{author}{P.~R. {Shapiro}}, and \bibinfo{author}{Y.~{Mao}},
  \bibinfo{journal}{\mnras} \bibinfo{volume}{\textbf{433}},
  \bibinfo{pages}{2900} (\bibinfo{date}{Aug. 2013}), \eprint{1304.6411}.
\bibitem{Mao13_reion_fNL}
\bibinfo{author}{Y.~{Mao}}, \bibinfo{author}{A.~{D'Aloisio}},
  \bibinfo{author}{J.~{Zhang}}, and \bibinfo{author}{P.~R. {Shapiro}},
  \bibinfo{journal}{\prd} \bibinfo{volume}{\textbf{88}}(8),
  \bibinfo{pages}{081303}, \bibinfo{eid}{081303} (\bibinfo{date}{Oct. 2013}),
  \eprint{1305.0313}.
\bibitem{1992PhR...215..203M}
\bibinfo{author}{V.~F. {Mukhanov}}, \bibinfo{author}{H.~A. {Feldman}}, and
  \bibinfo{author}{R.~H. {Brandenberger}}, \bibinfo{journal}{\physrep}
  \bibinfo{volume}{\textbf{215}}, \bibinfo{pages}{203} (\bibinfo{date}{Jun.
  1992}).
\bibitem{Jeong12}
\bibinfo{author}{D.~{Jeong}}, \bibinfo{author}{F.~{Schmidt}}, and
  \bibinfo{author}{C.~M. {Hirata}}, \bibinfo{journal}{\prd}
  \bibinfo{volume}{\textbf{85}}(2), \bibinfo{pages}{023504},
  \bibinfo{eid}{023504} (\bibinfo{date}{Jan. 2012}), \eprint{1107.5427}.
\bibitem{2006agna.book.....O}
\bibinfo{author}{D.~E. {Osterbrock}} and \bibinfo{author}{G.~J. {Ferland}},
  \bibinfo{title}{\emph{{Astrophysics of gaseous nebulae and active galactic
  nuclei}}} (\bibinfo{publisher}{Sausalito, CA:
  University Science Books}, \bibinfo{year}{2006}).
\bibitem{2004NewA....9..137W}
\bibinfo{author}{J.~W. {Wadsley}}, \bibinfo{author}{J.~{Stadel}}, and
  \bibinfo{author}{T.~{Quinn}}, \bibinfo{journal}{New Astronomy}
  \bibinfo{volume}{\textbf{9}}, \bibinfo{pages}{137} (\bibinfo{date}{Feb.
  2004}).
\bibitem{Viel_CosmoPars_Lya_2006}
\bibinfo{author}{M.~{Viel}} and \bibinfo{author}{M.~G. {Haehnelt}},
  \bibinfo{journal}{\mnras} \bibinfo{volume}{\textbf{365}},
  \bibinfo{pages}{231} (\bibinfo{date}{Jan. 2006}), \eprint{astro-ph/0508177}.

\end{thebibliography}

\appendix 

\section{Once more with gravity}\label{sec:adding-gravity}

The plan for this Appendix to regenerate equation
\eqref{eq:fLL-boltzmann} but now including peculiar velocity and
gravitational inhomogeneities in accordance with general
relativity. In fact, all the effects turn out to be minor on scales of
interest: dimensionless perturbations
to the metric $\phi$ are going to be small compared to the
dimensionless perturbations $\tdelta$ to the matter except on scales
comparable to or larger than the horizon:
\begin{equation}
\phi(\vec{k}) \sim \frac{3 H_0^2 \Omega_{m,0} (1+z)}{2 c^2 k^2}
\tdelta(\vec{k}) \approx \frac{4.2\,\mathrm{Gpc}^{-2}}{k^2} \tdelta(\vec{k})\textrm{,}
\end{equation}
where, as in the main text, $k$ is the comoving wavenumber. If you are
convinced by this argument, there is no need to read any further.

On the other hand, factors arising from spectral integrations could
outweigh the scale contrast and make the effects relevant. To be
sure either way one needs to press ahead with the calculation.
I work in conformal Newtonian gauge to make the geodesic equations
relatively simple, but will briefly discuss the effect of gauge changes
at the end of this Appendix.  The formal derivation starts with a
suitably perturbed flat Friedmann-Robertson-Walker (FRW) universe
described by the metric \cite{1992PhR...215..203M},
\begin{equation}
\dd s^2 = -(1+2\psi) c^2 \dd t^2 + a(t)^2 (1-2\phi) \dd \vec{x}^2\textrm{,}
\end{equation}
where $t$ is coordinate time, $\psi$ and $\phi$ are the scalar
potentials, and $\vec{x}$ are the comoving position coordinates.

Consider a photon with wavevector $k_{\mu}$ traveling through this
perturbed metric. The null condition $k_{\mu} k^{\mu}=0$ implies that
(working throughout at first order in the potentials)
\begin{equation}
\frac{\dd \vec{x}}{\dd t}=\frac{c}{a}(1+\psi+\phi)\vec{n}\textrm{,}
\end{equation}
where $\vec{n}$ is the unit vector in the spatial propagation
direction. The observed frequency of the photon in the coordinate
frame is $\nu = c k^{0}(1+\psi)$; combining this with the geodesic
equation for $k^{\mu}$ one finds that
\begin{equation}
\frac{\dd \nu}{\dd t} = \nu \left(-H - \frac{c}{a} \vec{n}
  \cdot \nabla \psi + \dot{\phi} \right)\textrm{.}\label{eq:redshift-gr}
\end{equation}
I will assume that on large scales all absorber and emitter streaming
velocities follow that of the pressureless dark matter. The tangent
4-vector $u^{\mu}$ with $u_{\mu} u^{\mu}=-1$ can be related to the
peculiar velocity $\vec{v}$ as
\begin{equation}
u^{\mu} = c^{-1} \left(\begin{array}{c}
1-\psi \\
\vec{v}/a\\
\end{array}\right)\textrm{,}
\end{equation}
where I have used a new assumption that $|\vec{v}|/c$ is small (the same order
as the potentials).  The frequency of the photon as seen by an
absorber is then
\begin{equation}
\nu' \equiv -u^{\mu} k_{\mu} \approx \nu \left[1-
  \vec{n}\cdot \frac{\vec{v}}{c} \right]\textrm{,}\label{eq:frequency-shift}
\end{equation}
again at first order. Finally, the physical 3-volume of a fixed
$\vec{x}$ coordinate patch is proportional to $a^3(1-3 \phi)$.

To define what is meant by an equilibrium solution to the Boltzmann
equation in the relativistic setting, consider the rate of change
$\dot{f}$ of the distribution function along a dark matter
worldline. We have:
\begin{equation}
\dot{f} = c u^{\mu} f_{,\mu} \approx (1-\psi)
\frac{\partial{f}}{\partial{t}} + \frac{\vec{v} \cdot \nabla f}{a}\textrm{,}
\end{equation}
and since the gradient term is overall second order, we can again
adopt the simple assumption that $\partial f/ \partial t=0$ to obtain
a well-defined equilibrium at first order, independent of gauge.

Putting this together, the underlying number density $f$ of photons
satisfies the collisional Boltzmann equation,
\begin{equation}
\frac{c}{a}
(1+\phi+\psi) (\vec{n} \cdot
\nabla) f + \frac{\partial f}{\partial \nu} \frac{\dd
  \nu}{\dd t} + \left(\frac{\dd \ln \Delta V\Delta \nu}{\dd t} \right)f = C_{\nu'}[f]\textrm{,}\label{eq:Boltzmann-full}
\end{equation}
where $C_{\nu'}[f]$ is calculated according to equation
\eqref{eq:collision-terms} as before, but evaluated at the
Doppler-shifted frequency $\nu'$ according to equation
\eqref{eq:frequency-shift}; $\nabla$ again means the ordinary
derivative with respect to comoving spatial coordinates $\vec{x}$,
and I have taken $\partial f/ \partial t=0$ as explained above. The
quantity $\Delta V \Delta \nu$ appears because $f$ is expressed in
photons per unit physical volume per unit frequency. A bundle of
photons which occupies a volume $\Delta V \Delta \nu$ in this space at
one moment will occupy a different volume at the next.  The evolution
of $\Delta V$ can be calculated by setting up an initially cubic
volume with edges $\vec{\Delta x}^{(\parallel)}$, $\vec{\Delta
  x}^{\perp1}$ and $\vec{\Delta x}^{\perp2}$ such that $\vec{\Delta
  x}^{\parallel}$ is parallel to $\nabla(\psi + \phi)$ and the others
are perpendicular. Then
\begin{align}
\Delta V & =  a^3(1-3 \phi) \Delta
  \vec{x}^{\parallel} \cdot \left( \Delta \vec{x}^{\perp1}
  \times \vec{\Delta x}^{\perp2}\right) \\
\Rightarrow \frac{\dd \Delta V}{\dd t} &= \frac{\dd \Delta
  \vec{x}^{\parallel}}{\dd t} \cdot \left( \Delta \vec{x}^{\perp1}
  \times \vec{\Delta x}^{\perp2}\right)a^3 (1-3 \phi) + 3 \Delta V
(H-\dot{\phi}) \nonumber \\
&= \Delta V \left(\frac{c}{a} \vec{n} \cdot \nabla (\psi + \phi) + 3H
  - 3\dot{\phi}\right)
\end{align}
at first order, with other terms canceling from the choice of
$\vec{\Delta x}$ vectors. Along with the frequency factor, which
follows immediately from equation \eqref{eq:redshift-gr}, the overall
volume term then reads
\begin{equation}
\frac{\dd \ln \Delta V \Delta \nu}{\dd t} = 2(H-\dot{\phi}) + \frac{c}{a}\vec{n}
\cdot \nabla \phi\textrm{.}\label{eq:phase-volume-gr}
\end{equation}

To link the gravitational effects to the perturbed density field
$\rho$, we require the Einstein equations.  The energy-momentum tensor
for the pressureless, matter-dominated universe is 
\begin{equation}
T_{\mu\nu} = \rho_0(1+\delta_{\rho}) u_{\mu} u_{\nu}
\end{equation}
We will work at sufficiently high redshift that we can take
$\Lambda=0$.  The zero-order Einstein equations recover the Friedmann
and acceleration equations for the pressureless fluid universe; the
linear-order equations reduce to
\begin{align}
\phi & = \psi \\
c^2 \nabla( \dot{\phi}+ H\phi) & = -4 \pi G \rho_0 a v(\vec{x}) \\
c^2 \nabla^2 \phi & = 4 \pi G \rho_0 a^2 \delta_{\rho} + 3 a^2 H
\left(\dot{\phi} + H 
\phi\right) \\
0 &= \ddot{\phi} + 4H  \dot{\phi}\label{eq:evolution} 
\end{align}
Equation \eqref{eq:evolution} is solved by $\dot{\phi}=0$ (the
potential is frozen, which corresponds to putting the matter
perturbation $\delta_{\rho}$ in the growing mode);
substituting also the zero-order Friedmann equation $3H^2 / c^2 = 8
\pi G \rho$ we have
\begin{align}
(-2c^2\nabla^2 + 6 a^2 H^2)\phi &= -3 a^2H^2 \delta_{\rho}\label{eq:einstein-potential} \\
\textrm{and } 3 a H \vec{v} &= - 2 c^2 \nabla \phi\textrm{.}\label{eq:einstein-velocity}
\end{align}

Let us now follow exactly the same procedure as in Sections
\ref{sec:nonrel-radiative-transfer} and \ref{sec:linearisation} to
obtain our previous approximation but with the relativistic terms
present. Multiply equation \eqref{eq:Boltzmann-full} by $\sigHI$ and
integrate over all frequencies; then, comparing against equation
\eqref{eq:fLL-boltzmann} at linear order, only two extra terms
survive. In particular, the perturbation to $\dd \vec{x}/\dd t$ is
irrelevant because there are no spatial gradients in the background.
Of the two remaining terms, first consider how the peculiar velocities
induce an extra term on the right hand side:
\begin{align}
\int C_{\nu'}[f] \sigHI \dd \nu & \approx \int C_{\nu}[f] \sigHI \dd \nu
- \frac{\vec{n}\cdot\vec{v}}{c} \alpha_v H \fLL \textrm{,} \\
\textrm{where } \alpha_v & = \frac{1}{H \fLL} \int \frac{\partial C_{\nu}[f]}{\partial
  \ln \nu} \sigHI \dd \nu \textrm{;}\label{eq:alpha-v}
\end{align}
the first term is the same as in our original calculation. The
integral in the second term need be evaluated only at zero-order, for
which we can use the background (zero-order) Boltzmann equation
\eqref{eq:Boltzmann-full} in the form
\begin{equation}
\frac{1}{H}\frac{\partial C_{\nu}[f_0]}{\partial \ln \nu} = 2 \frac{\partial
  f_0}{\partial \ln \nu} -  \frac{\partial^2 f_0}{\partial (\ln \nu)^2}\textrm{.}
\end{equation}
With the above, again using the $z=2.3$ spectrum from Ref. \cite{HM12}, I
obtain $\alpha_v = -12.0$.

The only other remaining gravitational term is the gradient term in
\eqref{eq:redshift-gr} and \eqref{eq:phase-volume-gr}; this and the
velocity term discussed above appear in the effective opacity which
now reads
\begin{equation}
  \kappat =  \sigHIbar \nHI + \kappaclumpbar + \frac{H}{c} \left( \alpha_z
    + 2 + \frac{\vec{n} \cdot \vec{v}}{c} \alpha_v \right) +  \frac{\alpha_z+1}{a} \vec{n}\cdot \vec{\nabla}\phi \textrm{.}
\end{equation}

Using the Einstein constraint equation in the form
\eqref{eq:einstein-velocity} we can update our
expression \eqref{eq:delta-kappat-no-grav} for the fractional
variations in $\kappat$:
\begin{align}
\tdelta_{\kappat} & = \betaHI \tdeltaHI +  \betaclump
\tdelta_{\kappaclumpbar} + \beta_{\phi} \frac{i\vec{n} \cdot
\vec{k} \phi}{a \kappatb}\textrm{,}\label{eq:delta-kappat-full} \\
\textrm{ where } \beta_{\phi} & = \alpha_z+1 - \frac{2}{3} \alpha_v
\approx 10.6 \textrm{.}\label{eq:beta-phi}
\end{align}
With this updated definition, equation
\eqref{eq:sln-with-angular-deps} remains valid. Integrating
\eqref{eq:sln-with-angular-deps} over $\vec{n}$ to obtain the solution
for $\tdelta_{\Gamma}$ is slightly more involved because angular
dependence now appears on the numerator as well as denominator; I obtain
\begin{align}
\tdelta_{\Gamma} = -\beta_{\phi} \phi
+ \left[(1-\betaHI \betarec)\tdelta_j - \betaHI(1-\betarec)
  \tdelta_{\nHI} \right. \nonumber \\
\hspace{2cm} \left. -\betaclump \tdelta_{\kappaclumpbar} + \betaHI \betarec \tdelta_{\Gamma} + \beta_{\phi} \phi\right]S(k)\textrm{.}
\end{align}

\begin{figure}
\includegraphics[width=0.5\textwidth]{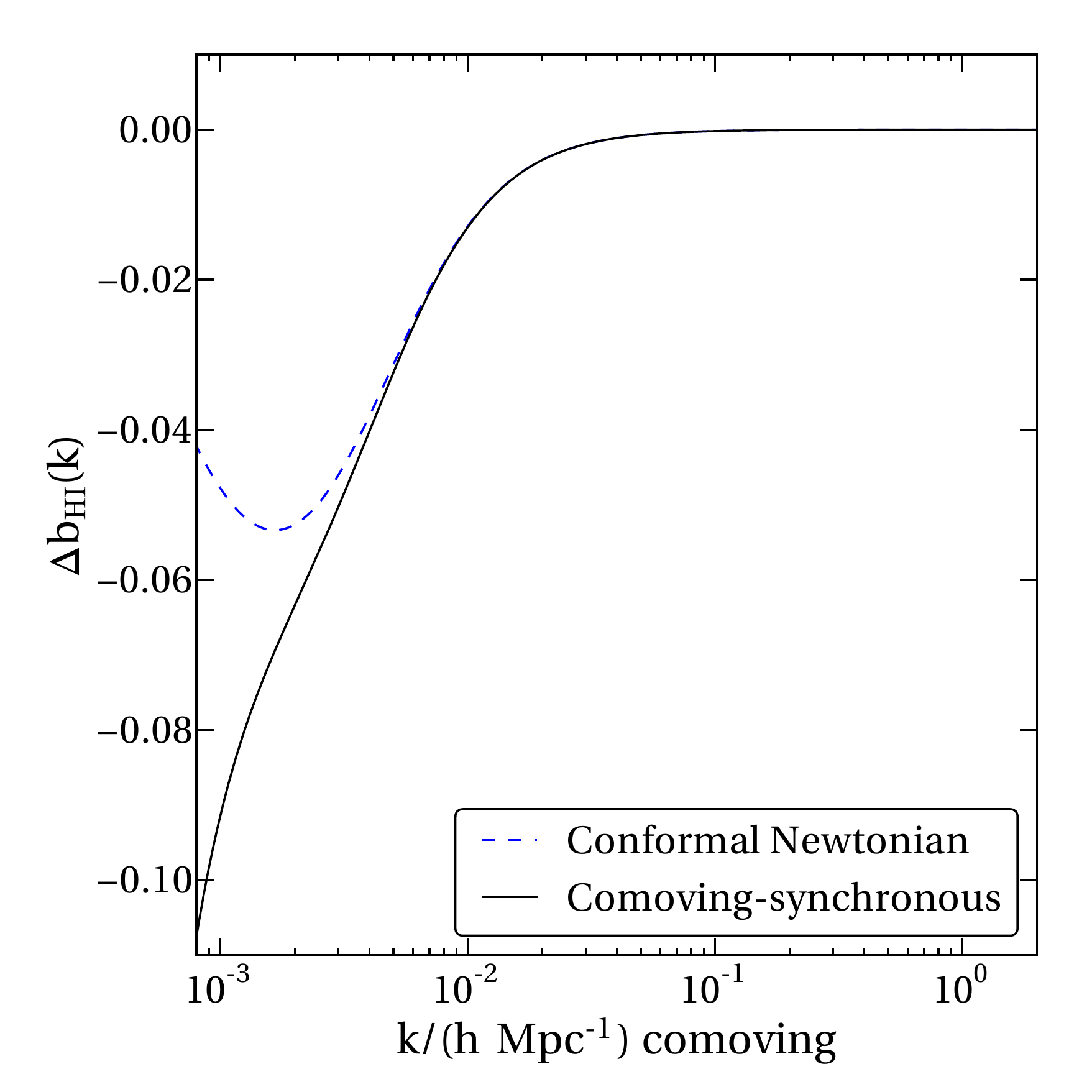}
\caption{The change in the \hi bias arising from the velocity and
  potential terms in the radiative transfer. The solid and dashed
  lines show the conformal-Newtonian \eqref{eq:grav-bias} and
  comoving-synchronous expressions respectively.  The effects are
  extremely minor as anticipated.}\label{fig:delta-bHI}
\end{figure}

The relation between local \hi density and radiation fluctuations is
still specified by equation \eqref{eq:gamma-to-nHI}, which allows us
to find the solution for the \hi fluctuations:
\begin{equation}
\hspace{0.8cm}
\tdelta_{\nHI} = \frac{\tdeltau + \beta_{\phi}\, \phi(k) - \left[\,\tdelta_{j,\mathrm{eff}} + \beta_{\phi}\,
  \phi(k)\right]S(k)}{1-\betaHI S(k)}\textrm{.}\label{eq:delta-nHI-relativistic}
\end{equation}
where $\tdeltau$ is the \hi density fluctuations in the absence of
radiation inhomogeneities. Writing the difference between equation
\eqref{eq:delta-nHI-relativistic} and \eqref{eq:delta-nHI-main} as
$\Delta \tdelta_{\nHI}$, one can define the change to the \hi bias from
the gravitational and Doppler effects:
\begin{equation}
\Delta \bHI = \frac{\Delta \tdelta_{\nHI}}{ \tdelta_{\rho}} = \beta_{\phi}
\frac{ 1-S(k)}{1-\betaHI S(k)}\frac{-3 a^2 H^2 }{2 c^2 k^2 + 6 a^2 H^2}\textrm{,}\label{eq:grav-bias}
\end{equation}
where I have made use of equation \eqref{eq:einstein-potential}. This
function is plotted in Figure~\ref{fig:delta-bHI} (dashed line), where
it can be seen that even on gigaparsec scales it reaches a maximum
shift of around $-0.05$, a tiny change in the bias (compare to Figure
\ref{fig:bk}). Note that the shot-noise component is unaffected.

It may be more natural to think of the bias on large scales in another
gauge -- it is more plausible, in particular, to imagine that $\bu$
and $b_j$ are scale-invariant in the comoving-synchronous than in the
conformal-Newtonian gauge \cite{Jeong12}.  (Ultimately one ought to
derive gauge-invariant observables, but for the reasons outlined in
the conclusions, that is beyond the scope of the current work.)  The
gauge transformation consists of a small coordinate transformation
$(t,\vec{x}) \to (t+T,\vec{x}+\vec{X})$; to reach the
comoving-synchronous gauge one uses the freedom to eliminate the
peculiar velocities in the coordinate frame. Following this through
for any quantity $Z$, assuming $Z_0 \propto
a^{-q}$, one finds that
\begin{equation}
\delta^N_{Z} = \delta^S_{Z}- 2q \phi/3\textrm{,}\label{eq:sync-to-newt}
\end{equation}
where superscript $N$ and $S$ stand for perturbations in
conformal-Newtonian and comoving-synchronous gauges
respectively. Equation \eqref{eq:sync-to-newt} can be used to
transform the conformal-Newtonian expression
\eqref{eq:delta-nHI-relativistic} into the synchronous equivalent,
giving
\begin{equation}
\tdeltaHI^S = \frac{\tdeltau^S + \beta_{\phi} \phi - \left[\tdelta_j^S +
  \beta_\phi\phi + \frac{2}{3}(\betaHI \qHI - \qjeff)\phi\right]
S(k)}{1-\betaHI S(k)}\textrm{.}
\end{equation}
Then, in the synchronous gauge, we have
\begin{equation}
\Delta \bHI^S =
\frac{-3 a^2 H^2 }{2 c^2 k^2 }\frac{ \beta_{\phi}-\left[\beta_{\phi} +\frac{2}{3}(\betaHI \qHI -
    \qjeff)\right] S(k) }{1-\betaHI S(k)}\textrm{,}\label{eq:grav-bias-syn}
\end{equation}
where $\Delta \bHI^S\equiv \Delta \tdelta^S_{\nHI} / \tdelta^S_{\rho}$,
and the result has been obtained using the relation between Newtonian
potential and synchronous-gauge density,
\begin{equation}
\phi = \frac{-3 a^2H^2}{2 c^2 k^2} \tdelta^S\textrm{.}
\end{equation}

To gain a quantitative picture we must estimate $\qHI$ and $q_j$. Note
that for any quantity $Y$ composed of a linear sum of components, $Y=\sum_i
Y_i$, one has
\begin{align}
q_Y & = \frac{\dd \ln Y}{\dd \ln a} = \frac{1}{Y} \sum_i \frac{\dd
  Y_i}{\dd \ln a} = \sum_i \beta_i q_{Y_i}
\end{align}
where $\beta_i=Y_i/Y$.  It therefore follows from using the background
equilibrium \eqref{eq:background-equilibrium} that
\begin{align}
q_j &\approx \left[\betaHI (1-\beta_r) + \betaclump\right]\qHI - \frac{3}{2}
\left[ \beta_z + \beta_V \right] \approx 2.7 \textrm{;} \nonumber \\
\qjeff &\approx (1- \betaHI \betar) q_j +(\betaHI \betar -
\betaclump)\qHI \approx 2.3\textrm{.}
\end{align}
where I have used $\dd \ln H / \dd \ln a \approx -3/2$ and $\qHI\simeq
4.3$, the latter from Ref. \cite{Rudie13MFP}. 

Adopting these estimates, equation \eqref{eq:grav-bias-syn} is plotted
as a solid line in Figure \ref{fig:delta-bHI}. The changes are larger
than in the Newtonian gauge but still small. It is worth noting that
the synchronous gauge bias as derived above describes a different
universe -- it is not, in fact, related by a gauge transformation to
the Newtonian case. This follows because I have formulated both
descriptions assuming a constant large-scale bias as an input
distribution; this assumption implies something physically different
in the two different gauges. As previously stated, it is probably a
more appropriate assumption in the synchronous than in the Newtonian
gauge.

This, however, is a tangential question because the gravitational and
Doppler effects are tiny in both gauges. The original decision to drop
these terms is therefore shown to be strongly justified.

\section{An estimate of $\bu$}\label{sec:estimate-b_0}

\begin{figure}
\includegraphics[width=0.5\textwidth]{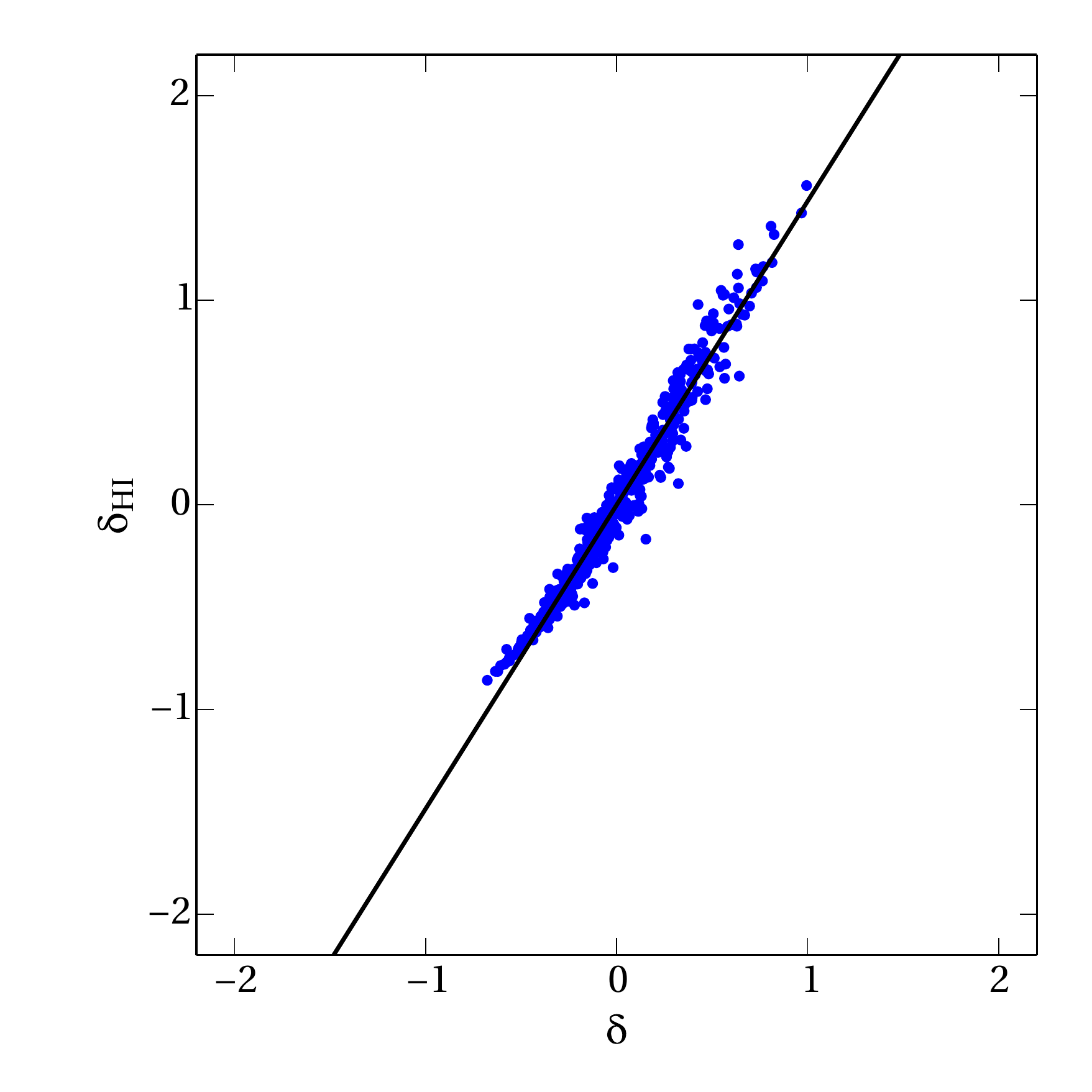}
\caption{The relationship between \hi overdensity and total matter
  overdensity in the IGM; each point represents the values averaged in
  a cube of side length $6\,\Mpc$ comoving at $z=2.3$. This derives
  from a simulation with uniform UV background. There is a very
  near-linear relationship between the two quantities, allowing the
  measurement of the slope (solid line), $\bu\approx 1.5$, which is a
  parameter entering the main calculation.}\label{fig:b0}
\end{figure}

To complete the calculation in the main text it was necessary to
specify a value of the bias $\bu$ of \hi in the absence of
inhomogeneous radiative effects. \updated{One can estimate this from the
photoionization equilibrium equations for a uniform field, coupled
with a description of the temperature-density relation for the
averaged IGM. Specifically, the uniform ionization equilibrium in the
limit that only a trace of neutral \hi survives specifies
\cite{2006agna.book.....O} that
\begin{equation}
\nHI \propto \frac{\alpha(T)\,\rho^2}{\Gamma_0}\textrm{,}
\end{equation}
where I have assumed the local electron and proton densities are both
proportional to the cosmic density $\rho$. Next write the
equation-of-state $T \propto \rho^{\gamma-1}$ and approximate
\cite{Black81clouds} $\alpha(T) \propto T^{-0.7}$; expanding both
$\rho$ and $\nHI$ in terms of their background values and
perturbations, one obtains
\begin{equation}
\bu = \frac{\deltau}{\delta_{\rho}} = 2-0.7(\gamma-1)\textrm{.}
\end{equation}
For a value \cite{HuiGnedin97_IGM_EoS} $\gamma=1.6$, this gives an
estimate of $\bu \simeq 1.6$. 

However, there is a slight inconsistency in the derivation above. The
recombination rate actually depends on the strictly local value of the
electron and proton densities, not on any linear-theory average on
large scales. Depending on how small-scale clustering reflects
large-scale density inhomogeneities, the assumption that the local
density scales with the environmental density may fail.} I therefore
also estimated $\bu$ directly from a cosmological simulation with
$256^3$ dark matter and $256^3$ gas particles in a $50\,\Mpc$-side
box. The code {\sc Gasoline} \cite{2004NewA....9..137W} implements
gravity, hydrodynamics, star formation feedback (which is likely of
minor importance here) and a uniform UV field, the values for which I
adopted from Ref.  \cite{HM12}.  Much more careful work has been
performed in simulating the forest by other authors
\cite{McDonald00,McDonald03,Viel_CosmoPars_Lya_2006} but they quote
statistics on the flux field, which is related to the \hi field by a
non-linear transformation and therefore does not directly tell us
$\bu$.

Taking the output at $z=2.3$, I interpolated the gas and dark matter
particles back onto a $256^3$ grid to obtain two 3D density maps, the
first of \hi and the second of total mass density with $\sim
0.2\,\Mpc$ resolution. To study the behaviour of the intergalactic
medium, I flagged all cells with dark matter density less than ten
times the cosmic mean. Only the flagged cells were subsequently used
to produce a degraded map with $8^3$ super-cells, in which the mean
dark matter and \hi density of the flagged sub-cells was recorded.

This allows us to see the large-scale relationship between IGM
overdensity and \hi (Figure \ref{fig:b0}). Each point represents an
IGM super-cell; the two axes correspond to dimensionless total mass
overdensity and \hi overdensity in the IGM, expressed as a fraction
according to equation \eqref{eq:define-delta}. The plots show a very
near-linear relationship between the total overdensity and the \hi
overdensity as expected. The slope of the line gives the bias, which
is found to be $\bu=1.48\approx 1.5$. \updated{This is in fair agreement with
the analytic estimate of $1.6$ given above, given the multitude of
uncertainties.}

I tested that this result is reasonably insensitive to the size of the
super-cells and the IGM threshold density. Repeating the exercise with
the IGM threshold at $\delta=5$, for instance, gives $\bu=1.45$; with
the original threshold but $16^3$ supercells, I retrieve $\bu=1.43$,
although the non-linearity in the relation starts to become more
prominent (as we are probing smaller scales) so the fit is less
meaningful. For the illustrative purposes of this paper, adopting
$\bu=1.5$ seems to pin down the large-scale relationship to a
sufficient accuracy.

\end{document}